\documentclass[epjc3]{svjour3}  

\usepackage{multirow}
\usepackage{diagbox}
\usepackage{subcaption}
\usepackage{tocloft}
\usepackage[countmax]{subfloat}
\usepackage{amssymb}
\usepackage{amsmath}
\usepackage[dvipsnames]{xcolor}
\usepackage{graphicx}
\usepackage{verbatim}
\usepackage{slashed}
\usepackage{hyperref}
\usepackage{makecell}
\usepackage{side}
\usepackage{xspace}
\usepackage[english]{babel}

\usepackage[right=2.5cm]{geometry} 

\usepackage{graphicx}
\usepackage[normalem]{ulem} 

\DeclareRobustCommand{\Sec}[1]{Sec.~\ref{#1}}

\DeclareRobustCommand{\Tab}[1]{Table~\ref{#1}}

\DeclareRobustCommand{\Fig}[1]{Fig.~\ref{#1}}

\DeclareRobustCommand{\Eq}[1]{Eq.~(\ref{#1})}

\newcommand{\reaction}{\texttt{EFT}\xspace reaction\xspace}
\newcommand{\ttbar}[0]{{t\bar{t}}}
\newcommand{\mtt}[0]{m_{t\bar{t}}}
\newcommand{\mt}{{m_{t}}}
\newcommand{\tmop}[1]{\rm #1}
\newcommand{\GeV}{\:\mathrm{GeV}}

\newcommand{\vb}[1]{{\texttt{#1}}}
\newcommand{\otq}{O_{tq}^{(8)}}
\newcommand{\ctq}{{c_{tq}^{(8)}}}

\newcommand{\ctqb}{{c_{tq8}}}
\newcommand{\otu}{O_{tu}^{(1)}}
\newcommand{\ctu}{{c_{tu}^{(1)}}}
\newcommand{\otd}{O_{td}^{(1)}}
\newcommand{\ctd}{{c_{td}^{(1)}}}

\newcommand{\ctg}{{c_{tG}}}
\newcommand{\dmt}{{\delta\mt}}
\newcommand{\TeVmt}{{\rm TeV}^{-2}}
\newcommand{\WC}{\mathbf{c}}

\newcommand{\PineAPPL}{\texttt{PineAPPL}\xspace}

\newcommand{\simunet}{\texttt{SIMUnet}\xspace}

\newcommand{\entry}[1]{\vspace{4mm}\noindent {\bf #1:}}
\newcommand{\entryi}[1]{\vspace{3mm}\noindent {\it #1:}}
\newcommand{\eftfile}{{EFT \vb{yaml} file\xspace}}
\newcommand{\equal}{{\mspace{1mu}=\mspace{1mu}}} 
\newcommand{\sigmaSMBSM}{\sigma(\WC)}
\newcommand{\sigmaSM}{\sigma(\WC\equal 0)}
\newcommand{\sigmaBSM}{\sigma(\WC)-\sigma(\WC\equal 0)}

\newcommand\YAMLcolonstyle{\color{darkgray}\bfseries}
\newcommand\YAMLkeystyle{\color{Sepia}\bfseries}
\newcommand\YAMLvaluestyle{\color{Violet}}
\definecolor{backcolour}{rgb}{0.95,0.95,0.92}
\definecolor{codegreen}{rgb}{0,0.6,0}

\makeatletter

\usepackage{listings}
\newcommand\language@yaml{yaml}

\expandafter\expandafter\expandafter\lstdefinelanguage
\expandafter{\language@yaml}
{
  keywords={true,false,null,y,n},
  backgroundcolor=\color{backcolour},   
  keywordstyle=\YAMLkeystyle\ttfamily,
  basicstyle=\YAMLkeystyle\ttfamily,   
  sensitive=false,
  comment=[l]{\#},
  morecomment=[s]{/*}{*/},
  commentstyle=\color{OliveGreen}\ttfamily,
  stringstyle=\YAMLvaluestyle\ttfamily,
  moredelim=[l][\color{orange}]{\&},
  moredelim=[l][\color{magenta}]{*},
  moredelim=**[il][\YAMLcolonstyle{:}\YAMLvaluestyle]{:}, 
  morestring=[b]',
  morestring=[b]",
  literate =    {---}{{\ProcessThreeDashes}}3
                {>}{{\textcolor{red}\textgreater}}1     
                {|}{{\textcolor{red}\textbar}}1 
                {\ -\ }{{\mdseries\ -\ }}3,
}

\lst@AddToHook{EveryLine}{\ifx\lst@language\languag@yaml\YAMLkeystyle\fi}
\makeatother
\newcommand\ProcessThreeDashes{\llap{\color{cyan}\mdseries-{-}-}}

\newcommand{\preprintnumber}{DESY-24-119}

\journalname{Eur. Phys. J. C}
\begin{document}

\title{A framework for simultaneous fit of QCD and BSM parameters with \textsc{xFitter}}

\author{Xiao-Min Shen\thanksref{e0,addr1,addr2,addr3} \and
  Simone Amoroso\thanksref{e1,addr1} \and
  Jun Gao\thanksref{e2,addr2,addr3} \and
  Katerina Lipka\thanksref{e3,addr1,addr4} \and
  Oleksandr Zenaiev\thanksref{e4,addr5}}

\institute{
  Deutsches Elektronen-Synchrotron DESY, Notkestr. 85, 22607 Hamburg, Germany\label{addr1}
  \and 
  {INPAC, Shanghai Key Laboratory for Particle Physics and Cosmology, School of Physics and Astronomy, Shanghai Jiao-Tong University, Shanghai 200240, China\label{addr2}}
  \and 
  {Key Laboratory for Particle Astrophysics and Cosmology (MOE), Shanghai 200240, China\label{addr3}}
  \and 
  {Fakult{\"a}t f{\"u}r Mathematik und Naturwissenschaften, Bergische Universit{\"a}t Wuppertal, Gau{\ss}strasse 20, D-42119 Wuppertal, Germany\label{addr4}}
  \and 
  {Institute for Theoretical Physics, Hamburg University, Luruper Chaussee 149, 22761 Hamburg, Germany\label{addr5}}
  }

\thankstext{e0}{e-mail: xiao-min.shen@desy.de}
\thankstext{e1}{e-mail: simone.amoroso@cern.ch}
\thankstext{e2}{e-mail: jung49@sjtu.edu.cn}
\thankstext{e3}{e-mail: katerina.lipka@desy.de}
\thankstext{e4}{e-mail: oleksandr.zenaiev@desy.de}

\date{Received: date / Accepted: date}
\maketitle
\begin{flushright}
\preprintnumber
\end{flushright}

\begin{abstract}
An extension of the \textsc{xFitter} open-source program for QCD analyses is presented, 
allowing for a polynomial parameterization of the dependence of physical observables on theoretical parameters.
This extension enables simultaneous determination of parton distribution functions (PDFs) and new physics parameters within the framework of the Standard Model Effective Field Theory (SMEFT).
The functionalities of the code are illustrated in a sensitivity study, where a simultaneous determination of the PDFs, top quark mass and the couplings of 
selected dimension-6 SMEFT operators is addressed using projections for measurements of top quark-antiquark pair production  at the High-Luminosity LHC.
The importance of considering all the correlations of the parton distributions, top quark mass
and the SMEFT parameters in a simultaneous QCD+SMEFT analysis is demonstrated.
This work serves as a new platform for simultaneous extraction of the PDFs and the SM/SMEFT parameters based on \textsc{xFitter}.
\end{abstract}

\setcounter{tocdepth}{1} 
\tableofcontents

\newpage

\section{Introduction}
Investigating physics at the highest energy scales is one of the primary objectives of the physics
program of the Large Hadron Collider (LHC) at CERN. 
Lack of direct evidence for new particles and interactions suggests that the scale of physics beyond the Standard Model (BSM) is higher than those probed directly at the LHC. 
However, the influence of fundamental new interactions could still show up indirectly through small deviations of experimental measurements 
from the SM predictions. 
The Standard Model Effective Field Theory (SMEFT)~\cite{Brivio:2017vri}
provides a systematic framework 
for incorporating these higher-energy effects into low-energy theories, 
allowing for investigations of new physics phenomena in a controlled and organized approach. 
The SMEFT extends the SM Lagrangian $\mathcal{L}_{SM}$
by a series of operators, $\mathcal{O}_i^{(d)}$ of dimensions $d>4$,
and effective couplings $c_i^{(d)}=C_i^{(d)}/\Lambda^{(d-4)}$ (Wilson coefficients),
suppressed by appropriate powers of
the energy scale $\Lambda$,  at which the EFT low-energy approximation loses validity.
Assuming lepton number conservation, the SMEFT Lagrangian may be written as
\begin{equation}
\mathcal{L}_{\mathrm{SMEFT}} = \mathcal{L_{\mathrm{SM}}} +\sum_i\frac{C^{(6)}_i}{\Lambda^2}\mathcal{O}_i^{(6)} + \cdots
~.
\label{eq:Lag-SMEFT}
\end{equation}
Constraints on these high-dimensional operators 
in the SMEFT framework  have been derived by different groups
using a variety of experimental data from the LHC.
These analyses benefit greatly from the high energies achieved at the LHC, 
due to energy-dependent factors which amplify the impact of  SMEFT operators on partonic cross sections.
These interpretations of the LHC data rely on a detailed knowledge of the proton structure, 
described in terms of Parton Distribution Functions (PDFs)~\cite{Gao:2017yyd}.
The PDFs cannot (yet) be calculated from the first principles and have to be determined using collider data itself, nowadays including also measurements from the LHC, and, in general, assuming no presence of BSM physics.
Full consideration of the correlation between the SMEFT operators and the proton PDFs would potentially modify the constraints on both, PDFs and the SMEFT couplings.
This correlation, however, has so far been disregarded in most but a few analyses~\cite{Greljo:2021kvv,CMS:2021yzl,Gao:2022srd,Shen:2024uop,Kassabov:2023hbm,Costantini:2024xae,Hammou:2023heg}.
{In Ref.~\cite{Greljo:2021kvv}, for example, interplay between PDFs and EFT effects are accessed for high-mass Drell-Yan processes at the LHC. 
 The necessity of joint PDF and EFT fit has also be studied using di-lepton
 invariant mass distribution and the forward-backward asymmetry in Drell-Yan processes in Refs.\cite{Hammou:2023heg,Anataichuk:2023elk},
 for top quark production in Refs.\cite{Gao:2022srd,Costantini:2024xae},
 and for jet production in Refs.\cite{CMS:2021yzl,Gao:2022srd}. The general conclusion of these investigations is that while the interplay between the PDFs and EFT effects remains moderate considering the current experimental precision, the combined SMEFT and PDF fit will be necessary for the interpretation of high-precision measurements at the High-Luminosity LHC (HL-LHC).  
}
This work presents a development of 
{a new public extension for simultaneous SMEFT and PDF fit}
within the open-source QCD analysis framework \textsc{xFitter}~\cite{Alekhin:2014irh,xFitterDevelopersTeam:2017xal,xFitter:2022zjb},
which has been widely used in experimental community for quantitative comparison of the data with theoretical predictions, determination of PDFs, fragmentation functions and SM parameters. 
Through its flexible and modular structure, \textsc{xFitter} provides an interface to a number of programs providing PDF evolution, heavy flavour schemes,
assumptions on the PDF parameterisation, and further theoretical inputs.
Besides actual PDF fits within HERA and the LHC collaborations, the \textsc{xFitter} framework has been used
for extractions of SM parameters as electroweak mixing angle~\cite{CMS-PAS-SMP-22-010}, the top quark mass $m_t$~\cite{Garzelli:2023rvx} or the strong coupling constant $\alpha_S(m_Z)$~\cite{ATLAS:2023lhg} and to study effects of physics beyond the SM~\cite{Anataichuk:2023elk}.
In Ref.~\cite{CMS:2021yzl} a simultaneous determination of $\alpha_S(m_Z)$, $m_t$ and Wilson coefficients
relevant for jet production~\cite{Gao:2013kp} was performed.
The interfaces used in these analyses to parameterise the cross section dependence on the studied parameters are, however, specific to the particular processes, parameters, and  models used.
In the present work, a new module of the \textsc{xFitter} code is presented, 
which can be used for the determination of generic SM parameters and EFT couplings, possibly with a simultaneous  extraction of the PDFs parameters. 
{The advantage of this work with respect to earlier studies, e.g. those of Ref.~\cite{Greljo:2021kvv}, is that the EFT corrections are stored in fast grids, which incorporate the full PDF dependence and accurate quadratic EFT corrections.}
%

The paper is organised as follows. 
In \Sec{sec:framework}, the fit framework is introduced,
together with a brief description of
the targets of SM/EFT parameters determination in \textsc{xFitter}.
As an illustration of the developed framework, a simultaneous SMEFT-PDF fit is performed, where the Wilson coefficients of selected dimension-6 SMEFT operators are determined together with the proton PDFs and the top quark pole mass, as discussed in~\Sec{sec:applications} and \Sec{sec:results}.
Conclusions are given in \Sec{sec:summary}.
The technical details of installation and using the code are given in Appendix~\ref{sec:install}.

\section{Framework on simultaneous fit of QCD and BSM}
\label{sec:framework}
The cross section for a generic process at the LHC, binned with respect to an observable $\alpha$, is factorised as:
\begin{equation}
\sigma^{(\alpha)} (\WC; \alpha_s, {\rm PDF}) = 
\sum_{a, b}
\int d x_1 d x_2 
 f_a (x_1) f_b (x_2) 
\hat{\sigma}^{(\alpha)}_{a b} (\WC; x_1, x_2, \alpha_s),
\label{eq:coll-fact}
\end{equation}
where 
$\WC=(c_1,c_2,\cdots)$ are BSM parameters to be fitted, $\alpha_s$ is the strong coupling constant, 
and $f_a(x)$ is the PDFs for a parton $a$,
and $\hat{\sigma}_{ab}^{(\alpha)}$ are the corresponding parton-level cross sections.
Details involving factorisation and renormalisation scales have been suppressed 
in \Eq{eq:coll-fact} for simplicity.
The BSM parameters,  $\alpha_s$, and the PDFs can be extracted (together or individually)
from the experimental data $D^{(\alpha)}$ by minimizing the profiled log-likelihood function
\begin{equation}
	\chi^2 (\WC,\alpha_s,{\rm PDF}) = \sum_{\alpha, \beta = 1}^{N_{{\rm pt}}}
    \big(\sigma^{(\alpha)} - D^{(\alpha)}\big) [{\rm cov}^{-
  1}]_{\alpha\beta}  \big(\sigma^{(\beta)} - D^{(\beta)}\big)\ ,
  \label{eq:chi2}
\end{equation}
where 
$N_{\rm pt}$ is the total number of data points used in the fit, and
${\rm cov}^{-1}$ is the inverse of the covariance matrix, incorporating 
both experimental and theoretical uncertainties. 
More details are given in e.g.~\cite{Gao:2017yyd}.
Determination of BSM parameters and PDFs using the $\chi^2$ method require 
both the precision and efficiency 
of the calculation of theoretical predictions $\sigma^{(\alpha)}$. 

Higher-order perturbative corrections 
to the SM partonic cross sections 
are essential for determination of various SM backgrounds and therefore drive the sensitivity of searches for BSM physics at the LHC.
Meanwhile, radiative corrections to the signal strength of BSM physics in
$\hat{\sigma}(\WC; \alpha_s)$
are equally important for accurate interpretation of the measurements and for constraining the BSM parameters.
The QCD corrections to BSM effects in terms of SMEFT have been developed for a long time in decays of $B$ mesons~\cite{Buchalla:1995vs}. 
Later, similar approaches have been applied to calculations of next-to-leading order (NLO) QCD corrections in decays of the top quark~\cite{Zhang:2008yn,Zhang:2010bm}, as well as in production of 
top quarks~\cite{Gao:2009rf,Gao:2011fx,Li:2011ek,Zhang:2011gh,Shao:2011wa} and jets~\cite{Gao:2011ha,Gao:2012qpa} at the LHC, induced by a subset of the SMEFT operators.
Recently, automated tools have been developed 
 for calculations of NLO QCD corrections to arbitrary hard processes including the contribution of  all dimension-6 SMEFT operators~\cite{Degrande:2020evl}.
These calculations and tools have been widely used in BSM searches at the LHC and lead to improved constraints on the Wilson coefficients of SMEFT operators. 
%

%
At the same time, additional complications arise in performing joint global analyses of QCD and the BSM effects at the LHC.
To incorporate the exact dependence of the theoretical predictions on the PDFs efficiently, fast-interpolation grid techniques~\cite{Carli:2010rw,Kluge:2006xs,Britzger:2012bs}, including PineAPPL grid~\cite{Carrazza:2020gss}, have been developed.
In general, the computational expensive 
parton-level cross sections $\hat{\sigma}_{a b}^{(\beta)}(\WC_0, x_1, x_2, \alpha_s)$ 
(for {\it fixed} BSM parameters $\WC_0$) 
is decomposed by order in $\alpha_s$, 
calculated once and stored in grids of parton flavors $a, b$, momentum fractions $x_1, x_2$, and bin $\beta$,
which can be later used in \Eq{eq:coll-fact}.
Again, complexity related to factorisation and renormalisation scales is suppressed here for simplicity.
Having incorporated dependence on coupling constants and renormalisation/factorisation scales, 
these grids, however, cannot handle also the cross section dependence on other SM parameters and the BSM parameters $\WC$.
To overcome this limitation, an implementation of dedicated analytic parameterisations, or development of interfaces to specific codes where 
the exact dependence can be implemented (either as tabulated data or using the grids) is necessary.
This is for example the case for the CIJET~\cite{Gao:2013kp} code, which allows to store and interpolate the contributions of certain SMEFT operators 
to the inclusive jet and dijet cross-sections via fastNLO-like grids~\cite{Kluge:2006xs,Britzger:2012bs}.
%

Fortunately, the dependence of observable on additional SM parameters or the BSM parameters $\WC$ can often be well described by a quadratic polynomial in the region concerned.
As an example, the cross sections of many processes depend on Wilson coefficients of dimension-6 SMEFT operators quadratically at NLO in QCD, if the renormalisation group running effects can be omitted, 
as is the case discussed in \Sec{sec:applications},
\begin{equation}
    \sigma(\WC) = \sigma_{\rm SM} + \sum_i \frac{C_i}{\Lambda^2} \sigma_i + \sum_{i,j} \frac{C_i C_j}{\Lambda^4} \sigma_{i,j} + \cdots ~.
\end{equation}
The assumption of quadratic dependence may also work well for the SM parameters. In the presented work, a specific case of the top quark mass $\mt$ is discussed. 
Even though the measured observable is in general a complicated function of $\mt$,
its dependence on $\mt$ in the region of interest (e.g for $\mt= 172.5\pm4.0$~GeV) can often be 
well approximated by a quadratic polynomial as
\begin{equation}
    \sigma(\mt\!=\!172.5\GeV+\dmt) = \sigma(\mt\!=\!172.5\GeV) + 
\sigma_{\dmt} \times\dmt + \sigma_{\dmt,\dmt}\times(\dmt)^2 \,,
\end{equation}
where $\sigma_{\dmt}$ and $\sigma_{\dmt,\dmt}$ are the linear and quadratic coefficients, respectively.
Upon this observation, 
a general solution for the aforementioned problem is discussed
in the present work and its implementation in \textsc{xFitter} in terms of a new \textsc{xFitter} module (so called \reaction) is documented.

%
To be specific, the \textsc{xFitter} \reaction  for a fit to a set of SM and/or BSM parameters $\WC=(c_1,c_2,\cdots)$, 
assuming that the contributions of $\WC$ to every fitted observable $\sigma^{(\alpha)}$
are well described by a polynomial of $\WC$ to quadratic\footnote{The limitation to quadratic interpolation is driven by a particular task to parametrise specific EFT couplings addressed, but the code can be extended to higher order polynomials.} terms, is
\begin{eqnarray}
  \sigma^{(\alpha)} (\WC; \alpha_s, {\rm PDF}) 
  & = & 
  \sigma_0^{(\alpha)}(\alpha_s, {\rm PDF}) + \sum_i c_i \sigma_i^{(\alpha)}(\alpha_s, {\rm PDF}) + \sum_{i \leqslant j} c_i c_j
  \sigma_{i j}^{(\alpha)}(\alpha_s, {\rm PDF}) \nonumber\\
  & = & \sigma_0^{(\alpha)} \left( 1 + \sum_i c_i K_i^{(\alpha)} 
  + \sum_{i\leqslant j} c_i c_j K_{i j}^{(\alpha)} \right). 
  \label{eq:quad-dep}
\end{eqnarray}
In the present work, the  definition of $\WC$ is extended to include further SM parameters.
The observables $\sigma^{(\alpha)}$
are not limited to cross sections and can be forward-backward asymmetries $A_{\rm{fb}}$ etc.
The superscript $\alpha$ runs over all bins of all used datasets.
Examples of SM/BSM parameters $\WC$
include SMEFT Wilson coefficients, and the
shift of the top quark mass $\delta m_t=m_t\!-\!172.5{\rm GeV}$, also studied in \Sec{sec:applications}.
%
%
Corresponding linear and quadratic contributions are denoted by $\sigma_i$, and $\sigma_{i j}$, respectively, with
$K_i=\sigma_i / \sigma_0$ and $K_{ij} = \sigma_{ij}/\sigma_0$ being the linear and quadratic $K$ factors.

Besides the quadratic dependence assumption in \Eq{eq:quad-dep}, 
the \textsc{xFitter} \reaction has barely limitation on 
the measured observables $\sigma^{(\alpha)}$ or fitted parameters $\WC$.
However, note that the linear/quadratic $K$ factors or corrections $\sigma_i, \sigma_{ij}$ have to be calculated separately prior to be used in the \reaction.
%
These input $K$ factors or theoretical predictions are usually calculated with a specific PDF set, 
neglecting their possible PDF dependence.
In order to study the correlation between the PDFs and SM/SMEFT parameters, 
the \textsc{xFitter} \reaction offers the option to read the linear/quadratic 
contributions $\sigma_i,\sigma_{ij}$ from fast interpolation grids.
To be specific, 
a set of fast interpolation grids corresponding to different choices of SM/BSM parameters \{$\sigma(\WC_1), \sigma(\WC_2), \cdots$\}
should to be fed to the \textsc{xFitter} \reaction, 
where a convolution of these grids with the PDFs is done, 
and 
the theoretical predictions for any $\WC$ are extracted by solving a system of linear equations. 
In this way, full PDF and SM/SMEFT dependence of the theoretical predictions is preserved during the fit.
The details about the installation and usage of the \textsc{xFitter} \reaction are given in Appendix~\ref{sec:install}, 
where possibilities to include
higher-power corrections beyond quadratic terms in \Eq{eq:quad-dep} are also discussed.

{In Refs.~\cite{Kassabov:2023hbm,Costantini:2024xae}, the \simunet program is presented, which is an open-source tool for simultaneous fits of EFT and PDFs.
It is built upon the NNPDF code, parameterizing PDFs using neutral networks.
The \simunet program makes use of the Monte Carlo replica method to propagate errors,
and is currently limited to linear EFT corrections \cite{Kassabov:2023hbm,Costantini:2024wby} via bin-by-bin $K$ factors.
Our work is based on the xFitter framework, which implies the $\chi^2$ minimisation according to \Eq{eq:chi2}, and provides various PDF parameterisations.
In our work, the EFT corrections are stored in fast grids, which incorporate the full PDF dependence and accurate quadratic EFT corrections.}

\section{Extraction of SMEFT couplings from $t\bar{t}$ production}
\label{sec:applications}
The functionality and capabilities of the new \reaction is illustrated by
a simultaneous determination of the PDFs,  $\mt$, 
and the couplings of four relevant SMEFT operators in \Eq{eq:ttbar-operator},
using an HL-LHC projection for a measurement of  the distribution of the invariant mass of the $t\bar t$ pair, $m_{t \bar t}$.
In this analysis, the HL-LHC pseudo-data are used
together with the combined HERA-I and II measurements of the inclusive deep-inelastic scattering (DIS) cross sections~\cite{H1:2015ubc}.  

\subsection{Theoretical predictions}
\label{sec:theo-pred}
Here, the selected dimension-6 SMEFT operators relevant for 
$\mtt$ distribution and the calculation of the SMEFT cross sections used to extract the $m_t$ and the Wilson coefficients are described. 
Furthermore, the PDF dependence of SMEFT linear/quadratic corrections is discussed, 
motivating the need to include the SMEFT corrections through fast-interpolation grids.

\subsubsection{SMEFT operators}
For BSM searches in the framework of SMEFT assuming lepton number conservation, 
the leading BSM contributions arise from dimension-6 SMEFT operators as in \Eq{eq:Lag-SMEFT}. 
In the present study, we consider the effect of the following operators relevant for $\ttbar$ production at the LHC:
\begin{eqnarray}
  O_{t u}^{1}&=& \sum_{i=1}^2 \left(\bar{t} \gamma_{\mu} t\right)\left( \bar{u}_{i} \gamma^{\mu} u_{i}\right)\,,\nonumber \\
  O_{t d}^{1}&=& \sum_{i=1}^3 \left(\bar{t} \gamma^{\mu} t\right)\left( \bar{d}_{i} \gamma_{\mu} d_{i}\right)\,, \nonumber\\
  O_{tG} &=& ig_s  (\bar{q}_{L 3} \tau^{\mu \nu} T^A t) 
  \tilde{\varphi} G_{\mu \nu}^A + \text{h.c.}\,, \label{eq:ttbar-operator}\\
  O_{tq}^8 &=&\sum_{i=1}^2 (\bar{q}_{L i} \gamma_{\mu} T^A q_{L,i})
               (\bar{t}  \gamma^{\mu} T^A t)\,. \nonumber
\end{eqnarray}
Here, 
$t$ is the right-handed top quark, 
$u_{i}, d_{i}$ are the right-handed quarks of the $i$-th generation,
$q_{L i}$ is the left-handed quark doublet of the $i$-th generation, 
$\varphi$ is the Higgs doublet, 
$G^A_{\mu \nu}$ is the gluon field strength tensor, and $g_s$ is the strong coupling. 
The associated Wilson coefficients $\ctu, \ctd, \ctg, \ctq$ are assumed to be real numbers.

The limits on the addressed Wilson coefficients have been extracted in various works,
with a few demonstrative examples listed below and summarised in the~\Tab{tab:WC-bound}.
The ATLAS collaboration \cite{ATLAS:2022xfj} has reported marginalised 95\% confidence level (CL) intervals
on $\ctg$ and $\ctq$, taking into account the linear EFT corrections at LO in QCD.
Further, there are also limits  extracted in the global fits. 
The SMEFiT collaboration~\cite{Celada:2024mcf}
has imposed constraints to  50  dimension-6 SMEFT operators
at NLO in QCD by fitting Higgs boson, diboson, and top quark production measurements at the LHC 
and electroweak precision observables from LEP and SLD.
Both linear and quadratic corrections in the 1/$\Lambda^2$ expansion are accounted for.
The 95\% CL limits for $\ctg$ and $\ctq$ are also extracted by the Fitmaker group~\cite{Ellis:2020unq},
by using the top quark, Higgs boson, diboson and electroweak measurements in a fit at leading order in QCD and considering linear EFT corrections only, not considering $\ctu$ and $\ctd$, since only weak constraints can be obtained with linear EFT contributions. 
The Wilson coefficients have also been constrained in the simultaneous SMEFT+PDF 
analysis in the CT18 global analysis framework
\cite{Gao:2022srd} with additional top quark and jet production measurements at the LHC and the Tevatron, resulting in the limits at 90\% CL, assuming $\ctu=\ctd$. 
Further results based on the data used in NNPDF4.0 global analysis~\cite{NNPDF:2021njg} are available, 
using SMEFT+PDF fit in the SIMUnet \cite{Kassabov:2023hbm,Costantini:2024xae} framework and considering only linear EFT corrections.

\begin{table}[h]
\centering
  \begin{tabular}{l|l|l|l|l}
    \hline
    & $\ctg$ [${\rm TeV}^{-2}$]& $\ctq$ [${\rm TeV}^{-2}$] & $\ctd$ [${\rm TeV}^{-2}$]& $\ctd$ [${\rm TeV}^{-2}$] 
    \\\hline
    ATLAS\cite{ATLAS:2022xfj} & $ [-0.68, 0.21]$  & $[-0.30, 0.36]$ &  N/A & N/A 
    \\\hline
    SMEFiT3.0\cite{Celada:2024mcf} & $[0.019, 0.180]$ & $[-0.467, 0.208]$ & $[-0.198, 0.186]$ & $[-0.159, 0.139]$
    \\\hline
    Fitmaker~\cite{Ellis:2020unq} & $[0.12, 0.6]$ & $[- 13, 2.2]$ & N/A & N/A 
    \\\hline
    CT18\cite{Gao:2022srd} & $- 0.10^{+ 0.26}_{- 0.30}$ & $- 0.8^{+ 2.58}_{- 2.38}$ & $0.14^{+
    0.61}_{- 0.97}$ & $=\ctu$ 
    \\\hline
  \end{tabular}
  \caption{
  Current limits on the Wilson coefficients obtained by 
  ATLAS, SMEFiT, Fitmaker, and CT18 collaborations.
  The results of ATLAS, SMEFiT, Fitmaker are the marginalised intervals correspond to 95\% CL, while the results of CT18 are at 90\% CL.
  }
  \label{tab:WC-bound}
\end{table}

%
\subsubsection{Calculations of the theoretical predictions}

The theoretical predictions for $t\bar t$ production 
in SM are calculated at NNLO in QCD using \vb{MATRIX}~\cite{Grazzini:2017mhc,Catani:2019hip} interfaced to \PineAPPL~\cite{Carrazza:2020gss}.
The nominal value of the top quark pole mass is set to $m_t$=172.5~GeV.
Predictions with $m_t$ values of {167.5, 170, 175, 177.5}~GeV are further generated to interpolate the $m_t$ dependence.
The renormalization and factorisation scales are set to 
$\mu_F = \mu_R = H_T/4$, with $H_T\equiv \sqrt{m_t^2+p_{T,t}^2}+\sqrt{m_t^2+p_{T,\bar{t}}^2}$.

The EFT corrections are calculated at NLO in QCD 
using Madgraph5\_aMC@NLO~\cite{Alwall:2014hca} together with the SMEFT@NLO~\cite{Degrande:2020evl} model,
using the same dynamical renormalisation and factorisation scales as in the SM predictions.
The \PineAPPL grids with various non-vanishing $\ctg, \ctq$, $\ctu$ and $\ctd$ values
are generated to account for both the linear and quadratic dependence.
In this work, we omit the renormalisation group running\footnote{Renormalisation group running effects of the dimension-6 operators 
has been studied at LO in QCD for top quark production and Higgs boson production in Refs.~\cite{Aoude:2022aro,Maltoni:2024dpn}, 
and at NLO in QCD for jet production in Ref.~\cite{CMS:2021yzl}.} of the Wilson coefficients,
as most of the other groups~
\cite{Greljo:2021kvv,Gao:2022srd,Kassabov:2023hbm,Costantini:2024xae}.
As a result, the theoretical predictions of the cross sections depend on these Wilson coefficients quadratically as in \Eq{eq:quad-dep},
such that these coefficients
can be fitted by using the \textsc{xFitter} \reaction.
Though the \reaction can deal with the crossing terms between different the Wilson coefficients,
these are not included in the present work for simplicity.
For instance, corrections proportional to $\ctg^2$ are included, but not those from $\ctg\cdot\ctq$.

Predictions for the DIS reduced cross section are obtained at NNLO in QCD through QCDNUM~\cite{Botje:2010ay}
using the Thorne-Roberts~\cite{Thorne:1997ga,Thorne:2006qt,Thorne:2012az} variable-flavor number scheme.
The SMEFT operators considered in this work do not contribute to the
DIS cross section at NLO in QCD.

\subsubsection{Sensitivity of $\mtt$ distributions to $\mt$ and SMEFT coefficients}
\label{sec:mt-WC-dep}
\begin{figure}
    \centering
\includegraphics[width=0.9\linewidth]{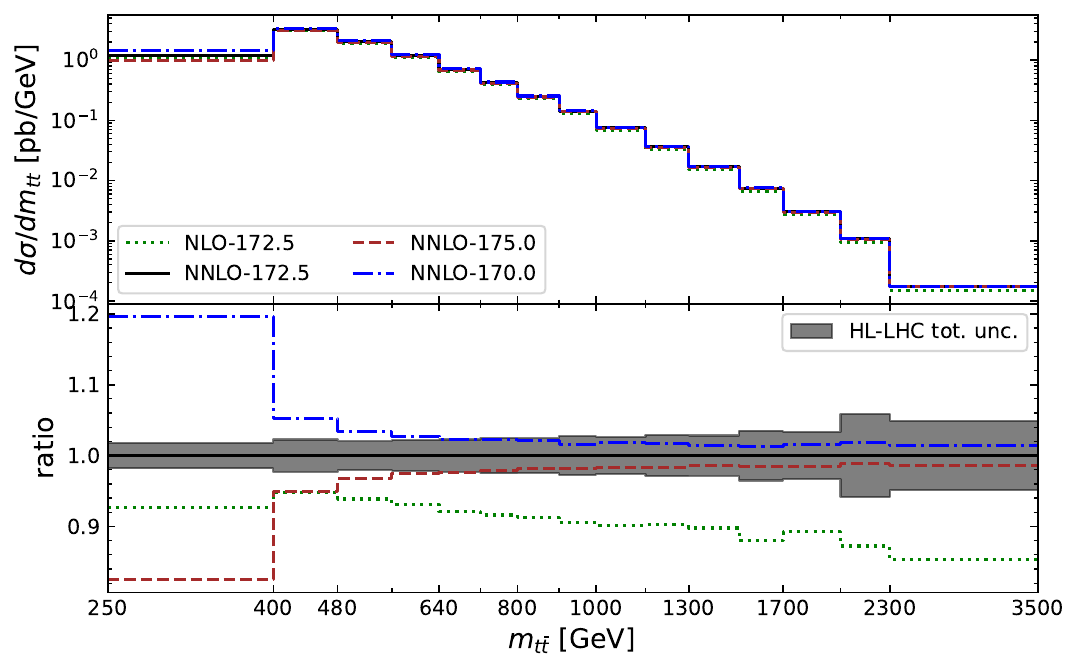}
    \caption{The NNLO QCD prediction for $\mtt$ distribution 
    for $t\bar t$ production at the LHC at $\sqrt{s}$=13~TeV, 
    assuming different values of $m_t$ (lines of different styles). The projection for the total uncertainty in a future HL-LHC measurement (shaded band) is also shown. 
    } 
    \label{fig:mt-dep}
\end{figure}

The values of $m_t$ and of the SMEFT coefficients are treated as free parameters in the fit.
In Fig.~\ref{fig:mt-dep}, the SM predictions for the $\mtt$ distributions at NNLO in QCD, are shown for the values of $m_t$ of 170, 172.5 and 175~GeV. 
The CT18NNLO PDF set~\cite{Hou:2019efy} is used.
The cross section exhibits largest sensitivity to the value of $m_t$ 
in the first bin of $\mtt$, close to the $2\cdot \mt$ threshold region.

\begin{figure}{t}
    \centering
\includegraphics[width=1.0\linewidth]{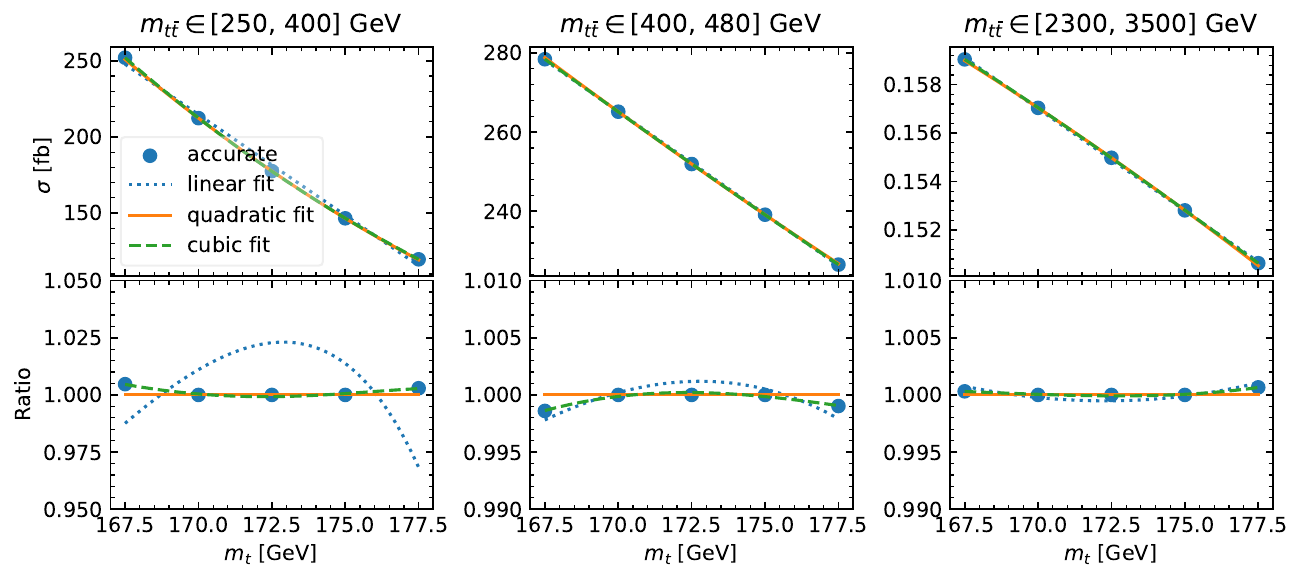}
    \caption{The absolute (upper panel) and normalised (lower panel) dependence of the $t\bar t$ cross section prediction in selected $\mtt$ bins on the value of $m_t$. The predictions obtained with $m_t$=167.5, 170, 172.5, 175 or 177.5~GeV, are shown by filled circles. The linear (dotted line), quadratic (solid line) and cubic (dashed line) polynomial interpolations are also shown.
    }
    \label{fig:mt-dep-poly}
\end{figure}

In \Fig{fig:mt-dep-poly}, the $m_t$ interpolation is illustrated,
considering the cross section predictions in the first, second and last $m_{\ttbar}$ bins. The exact predictions are shown together with the linear, quadratic and cubic polynomial fits.
While the the complete $m_t$ range is used for linear and cubic fits, it is narrowed to 170$<m_t<$175~GeV for quadratic fit to achieve higher precision around 172.5~GeV.
The corrections beyond quadratic power $\mathcal{O}(\mt \!-\! 172.5\GeV)^3$ are at per mill level for $\mt = 167.5$, $177.5~\GeV$.
\begin{figure}[h]
    \centering
\includegraphics[width=0.75\linewidth]{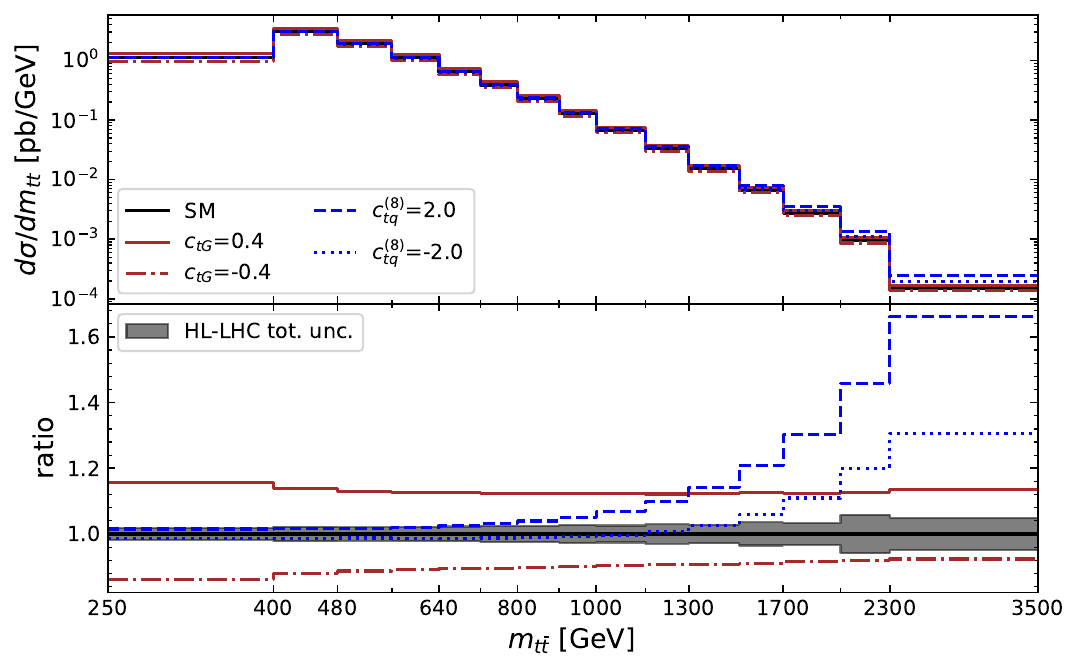}
 \includegraphics[width=0.75\linewidth]{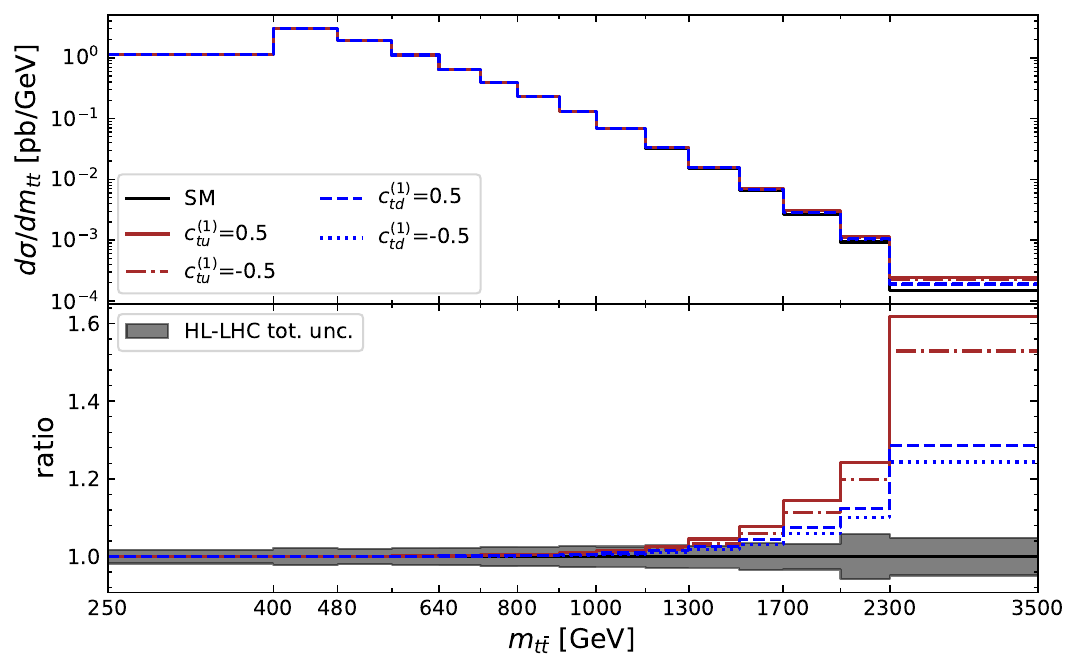}
    \caption{
    Predictions for $\mtt$ in SM and SMEFT at NLO in QCD, for different values of $\ctg$ and $\ctq$ (upper) or 
    $\ctu$ and $\ctd$ (lower),
    shown by lines of different styles. The ratios are shown in the lower panels.  
     In each SMEFT prediction, only one out of $\ctg$, $\ctu, \ctd$, and $\ctq$ is considered non-zero for clarity.
    All the Wilson coefficients are shown in unit of $\TeVmt$. The projection for the experimental uncertainty in HL-LHC scenario is shown as a shaded band.}
    \label{fig:SMEFT-pred}
\end{figure}
The missing higher power corrections are expected to be 
at $\sim 0.01\%$ level
for 170$<m_t<$ 175~GeV.
In the following, the $m_t$ dependence interpolated by the quadratic form is used within the \reaction as in \Eq{eq:quad-dep}.

The $\mtt$ distributions assuming only one non-vanishing SMEFT parameter
are shown in \Fig{fig:SMEFT-pred}, 
together with the projection of the total experimental uncertainty expected at the HL-LHC. While the $O_{tG}$ operator
primarily modifies the overall $t\bar t$ production cross-section,
the operators $\otq, \otu$ and $\otd$ mainly contribute to the cross section at high $m_\ttbar$. Also, for $|\ctq|, |\ctu|$ and $|\ctd|$ around or larger than $0.5~\TeVmt$, 
the quadratic corrections tend to dominate the linear ones
and shall not be omitted in the fit.

\begin{figure}[htbp]
\centering
  \includegraphics[width=1.0\textwidth]{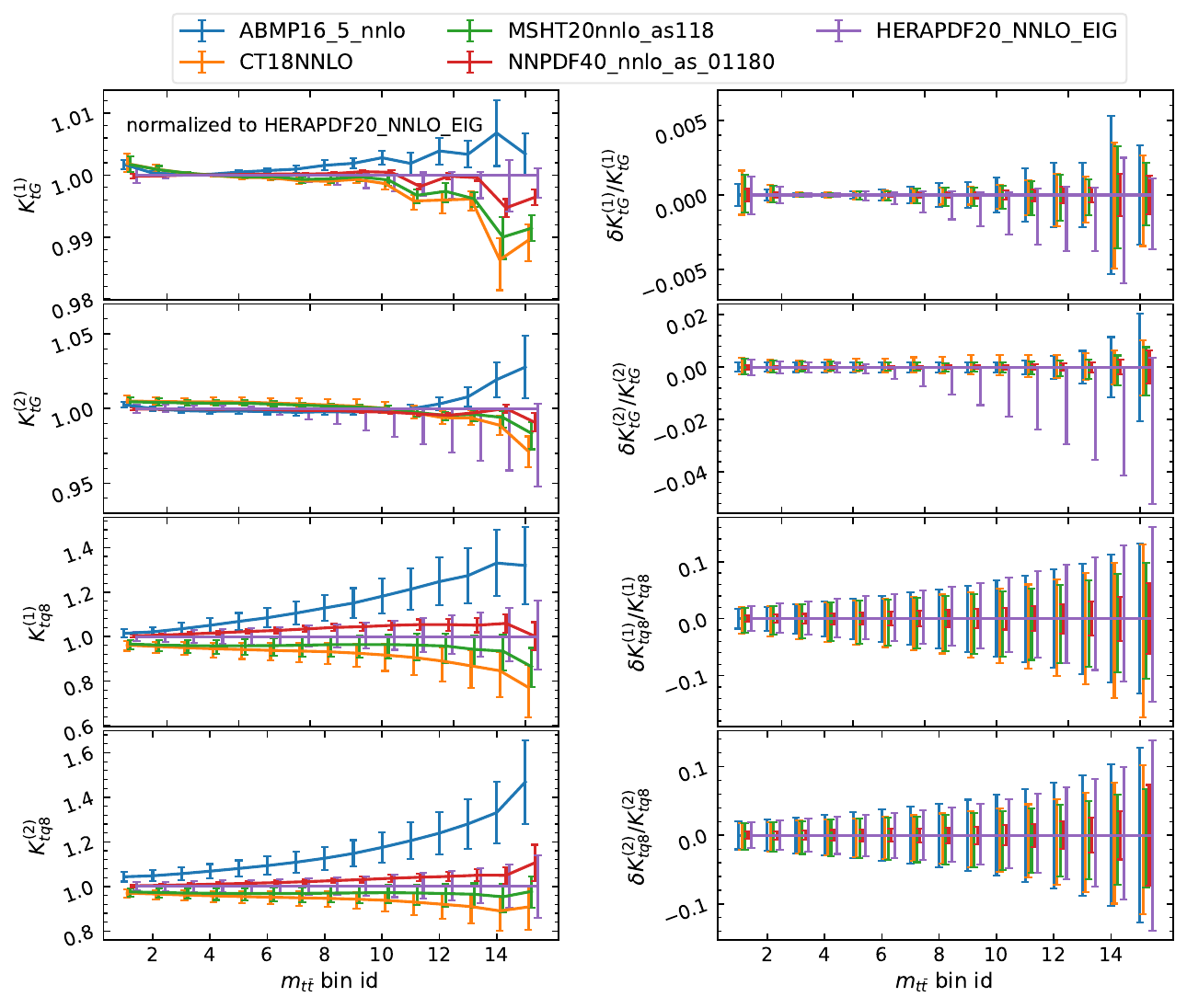}
	\caption{
The BSM $K$ factors calculated with ABMP16, CT18, MSHT20, NNPDF40 and HERAPDF20 PDFs, together with the 68\% CL PDF uncertainties in 15 $\mtt$ bins as in~\Fig{fig:SMEFT-pred}, used in Ref.~\cite{CMS:2021vhb}.
Shown are the linear and quadratic BSM $K$ factors for $\ctg$ and $\ctq$.
Results in the left (right) panel are normalised to the central member of HERAPDF20 (individual PDF).
	}
  \label{fig:BSM-KFactor}
\end{figure}

From the theoretical predictions calculated for various non-vanishing SMEFT parameters, 
the linear and quadratic BSM $K$ factors $K_i$ and $K_{ij}$ in \Eq{eq:quad-dep}
can be determined.  
In Fig.~\ref{fig:BSM-KFactor}, the BSM $K$ factors are presented. These are calculated by using 
ABMP16~\cite{Alekhin:2017kpj},
CT18~\cite{Hou:2019efy}, 
MSHT20~\cite{Bailey:2020ooq}, 
NNPDF4.0~\cite{NNPDF:2021njg} 
and HERAPDF2.0~\cite{Abramowicz:2015mha} PDF sets, 
considering the PDF uncertainties at 68\% CL.
The factors $K_{tq8}^{(1)}$ and $K_{tq8}^{(2)}$, for example, correspond to the linear
and quadratic BSM $K$ factors of the Wilson coefficient $\ctq$, respectively.
For $\ctg$, the PDF dependence of the BSM $K$ factors is small.
For $\ctq$, this dependence is much larger, as concluded from both, the individual PDF uncertainties and the differences in predictions obtained using different PDFs, reaching 20$\%$ at high $\mtt$. 
This observation motivates using the fast interpolation grids to store the contribution of $\ctq$ instead of using BSM $K$ factors 
derived with the fixed PDFs, in order to fully account for the correlation between PDFs and SMEFT parameters.

\subsection{The datasets}
\label{sec:data}
In the present work, the combined HERA inclusive DIS cross sections ~\cite{H1:2015ubc} are used together with a
projection for a future HL-LHC measurement of $t\bar t$ production as a function of $\mtt$.
The minimum $Q^2$ for the HERA DIS data is chosen to be $3.5~{\rm GeV}^2$.

The HL-LHC pseudo-data are simulated using the same binning as used in a CMS measurement~\cite{CMS:2021vhb} of differential cross sections of $t\bar{t}$ production in the semileptonic decay channel using the Run 2 data at $\sqrt{s}=$13~TeV. In particular, the same 15 bins of $m_{t\bar{t}}$ are used. 
The statistical uncertainties of Ref.~\cite{CMS:2021vhb} are scaled 
to their expectation values assuming an integrated luminosity of 3~ab$^{-1}$.
Following the prescription of Ref.~\cite{Azzi:2019yne}, the systematic uncertainties dominated by uncertainties in the b-jet calibrations and from theoretical modeling, are scaled down by a factor of two, 
while the uncertainty in the integrated luminosity of 1\% is used.
The nominal HL-LHC projection for the $\mtt$ distribution
corresponds to the SM theoretical predictions at NNLO in QCD 
with the value of $m_t$=172.5 GeV, and using HERAPDF2.0NNLO~\cite{Abramowicz:2015mha}.
Alternatively, a scenario is tested, where non-vanishing SMEFT contributions are present in the pseudo-data,  obtained by adding the EFT corrections to the SM cross sections. 
In all the presented studies, the SM cross sections are calculated at NNLO in QCD, while the EFT corrections are calculated at NLO in QCD and assuming $\ctg=-0.1 ~\TeVmt$ and 
$\ctq=1.0 ~\TeVmt$.


\subsection{PDF parameterisation}

The general strategy for a simultaneous QCD and BSM fit, introduced in \Sec{sec:framework}, is used. The QCD analysis is performed at NNLO, with the DGLAP~\cite{Gribov:1972ri,ALTARELLI1977298,CURCI198027,FURMANSKI1980437,Moch:2004pa} evolution implemented in QCDNUM~\cite{Botje:2010ay}
version 18-00-00.
{QED corrections can be included into QCD evolution in the xFitter framework, but are neglected in this work since their effects are small for the processes under consideration.
We also omit the impact of New Physics on the evolution of PDFs.
The scale of New Physics in the SMEFT framework are assumed to be well above the electroweak scale, so new particles will not change radiations of QCD partons or the evolution of PDFs. 
}
The values of heavy quark masses are set as $m_c$ = 1.43 GeV and $m_b$ = 4.5 GeV, while $m_t$ is a free parameter of the fit. 
The value of $\alpha_S(m_Z)$ is set to 0.118. 
The PDF parameterisation corresponds to the one used in HERAPDF2.0 analysis~\cite{Abramowicz:2015mha}. 
The parameterised PDFs are the gluon $xg(x)$, the valence quark $x u_{v}(x)$ and $xd_{v}(x)$ distributions, as well as $x\bar U(x)$ for the up- and $x\bar D(x)$ for the down-type antiquarks. At the starting scale of QCD evolution $Q_0^2$ = 1.9 GeV$^2$, the parameterisation form for a PDF $\mathrm{f}$ is
\begin{equation}
x f(x) = A_{f} x^{B_f}(1-x)^{C_f} (1 + D_f x + E_f x^2).
\label{PDFgeneralForm}
\end{equation}
For the gluon distribution,  an additional term of the
form $A'_g x^{B'_g} (1-x)^{C'_g}$ is subtracted, allowing for a negative gluon at low $x$.
The normalisation parameters $A_{u_{\rm v}}$, $A_{d_{\rm v}}$, and $A_g$ can be determined from the QCD sum rules.
The small-$x$ behaviour of the PDFs is driven by the $B$ parameters, whereas the $C$ parameters are responsible for the behaviour as $x \to 1$. The relations $x\bar U(x) = x\bar u(x)$,
$x\bar s(x) = f_s x\bar D(x)$, and $x\bar d(x) = (1-f_s) x\bar D(x)$ are assumed at the starting scale, 
with $x\bar u(x)$, $x\bar d(x)$, and $x\bar s(x)$ 
being the distributions for the up, down, and strange antiquarks, respectively. 
Here,
$f_s$ is the strangeness fraction, fixed to $f_s = 0.4$ as in the HERAPDF2.0 analysis~\cite{Abramowicz:2015mha}.
Further constraints 
$B_{\bar U} = B_{\bar D}$ 
and $A_{\bar U} = A_{\bar D}(1 - f_s)$ are imposed, so that 
the $x\bar u$ and $x\bar d$ distributions have the same normalisation as $x \to 0$. 
Note, that the \textsc{xFitter} package interfaced with the \reaction is not bound to any particular parameterisation form.

Simultaneous fits of PDFs, $m_t$, and SMEFT Wilson coefficients 
are performed with \textsc{xFitter} incorporating   
the \textsc{xFitter} \reaction based on \Eq{eq:quad-dep}.
The quadratic $m_t$ dependence of the cross section is adopted, as justified in 
\Sec{sec:mt-WC-dep}.
The PDF parameters, the value of $m_t$ and the Wilson coefficients are extracted through minimisation of $\chi^2$, introduced in \Sec{sec:framework}, applying a criterion of $\Delta\chi^2=1$. In the present study, only Hessian fit uncertainties are considered.

\section{Results of joint SMEFT-PDF fits}
\label{sec:results}

Two alternative scenarios are considered in the present analysis, with respect to the generation of the $\mtt$ pseudo-data.
An SM scenario, where the pseudo-data do not contain any BSM effects, and a SMEFT scenario, where the $\mtt$ pseudo-data include SMEFT contributions computed with $\ctg=-0.1~\TeVmt$ and $\ctq=1.0~\TeVmt$, considering both linear and quadratic corrections. Besides investigating  the correlations between the SM and SMEFT parameters, these scenarios can be used to hint to possible bias in the PDF extraction in case the data would indeed contain BSM physics, or, alternatively, test a BSM hypothesis in case no BSM physics is present.
Note that DIS data from HERA, included for the PDF constraints, has no sensitivity to the value of $m_t$ or the investigated Wilson coefficients. 
Given the illustrative character of the present analysis, 
no attempt was made to derive a complete and accurate set of parameter uncertainties.
In particular, while uncertainties from the input data  are derived using an Hessian approximation at the minimum, 
there was no attempt to estimate uncertainties from missing higher orders.

\subsection{SM scenario for pseudo-data}
The SM scenario mimics no BSM effects present in the $\mtt$ distribution. The $t\bar t$ pseudo-data are generated using the SM NNLO QCD prediction with $m_t = 172.5~{\rm GeV}$,
and using the HERAPDF2.0 NNLO PDF set. 
Two alternative fits are performed:
in a fit denoted as {\it SM-SM}, the PDFs and $m_t$ are fitted, while in a {\it SM-SMEFT} fit, the PDFs, $m_t$, and the coefficients $\ctg,\ctq,\ctu,\ctd$ of the operators in \Eq{eq:ttbar-operator} are obtained. The results are summarised in the Table~\ref{tab:fit-res-SM} and are discussed in the following.
\begin{table}[h]
\centering
\setlength\tabcolsep{5 pt}
  \begin{tabular}{l|c|c|c|c|c}
    \hline
     & $m_t$& $\ctg$ & $\ctq$ & $\ctu$ & $\ctd$\\
         & [GeV]& $[\TeVmt]$ & $[\TeVmt]$ & $[\TeVmt]$ & $[\TeVmt]$\\
    \hline
    generated & 172.5 & 0 & 0 & 0 & 0 \\
    \hline
    SM-SM fit & $172.53 \pm 0.24$ & - & - & - & -\\
    \hline
    SM-SMEFT fit & $172.51 \pm 0.37$ & $- 0.01 \pm 0.08$ & $0.07 \pm 0.62$ & $-
    0.01 \pm 0.42$ & $0.00 \pm 1.08$\\\hline
  \end{tabular}
  \caption{
  Results for $m_t$ and the Wilson coefficients
  obtained in the fits to HERA DIS and SM $\mtt$ pseudo-data in the SM-SM and SM-SMEFT scenarios.
  The values in the first row correspond to the input parameters used to generate the pseudo-data. 
  \label{tab:fit-res-SM}
  }
\end{table}

\entry{SM-SM fit}
The only difference between the present SM-SM fit and the original HERAPDF2.0 NNLO analysis~\cite{Abramowicz:2015mha}
is the inclusion of the $\ttbar$ pseudo-data. 
\begin{figure}[h]
\centering
  \includegraphics[width=0.7\textwidth]{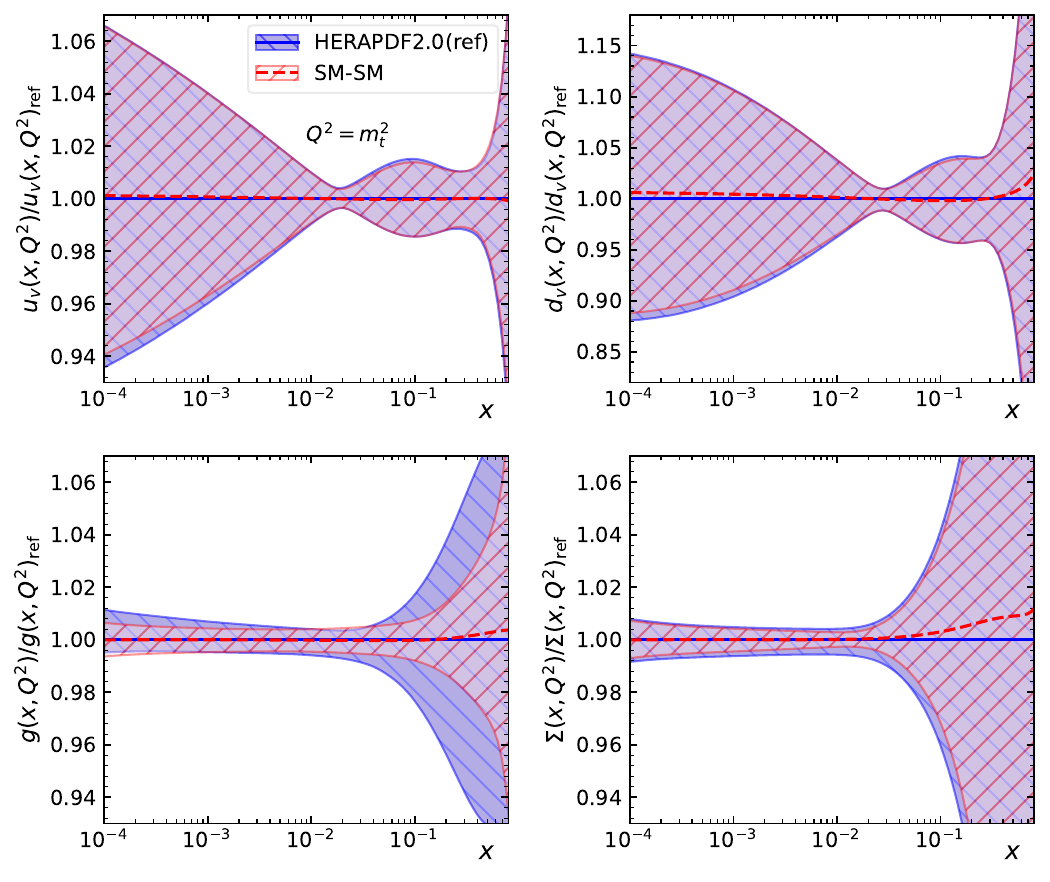}  
	\caption{
    The $u_v$, $d_v$, gluon and the sea PDFs obtained in the SM-SM fit (red shaded band), compared to the HERAPDF2.0NNLO PDFs~\cite{Abramowicz:2015mha} (blue shaded band), shown as a function of $x$ at the factorisation scale of $m_t$. The distributions are divided by HERAPDF2.0 and the SM-SM result is shown by a dashed line).
    }
  \label{fig:PDFs-SM-HERA}
\end{figure}
While HERA DIS is not sensitive to the value of $m_t$, the $t\bar t$ data impose an additional constraint on the gluon distribution at high $x$, since the $t\bar t$ pairs are predominantly produced in gluon-gluon fusion, and probe $m_t$ directly. 

The PDFs extracted in the present SM-SM fit are compared to the original HERAPDF2.0 PDFs in Fig.~\ref{fig:PDFs-SM-HERA}. 
The difference between the central PDFs is negligible considering the fit uncertainties.
The constraint on the gluon PDF improves at high $x$ in SM-SM fit, attributed to inclusion of the $\ttbar$ pseudo-data, while only minor impact on the quark PDFs is observed. The value of the top quark pole mass in SM-SM fit is obtained as $m_t=172.53 \pm 0.24$~GeV, consistent with the input value of 172.5 GeV. Again, only the Hessian fit uncertainty is considered, which is compatible to the current uncertainty in $\mt$ obtained in direct measurements at the LHC, e.g. the one of Ref.~\cite{CMS:2024yqd}. These results give confidence in the expectation that once the data contain only the SM contributions, the comprehensive QCD analysis would return the correct PDFs and the value of the SM parameters, in this example, $m_t$. While further uncertainties, e.g. arising from scale variations in the NNLO QCD predictions, will not change this conclusion, in an analysis based on real measurements these should be taken into account.

\entry{SM-SMEFT fit} In an alternative SM-SMEFT fit, an attempt is made to extract the BSM contributions from the data, where these are  not present. 
\begin{figure}[h]
\centering
  \includegraphics[width=0.7\textwidth]{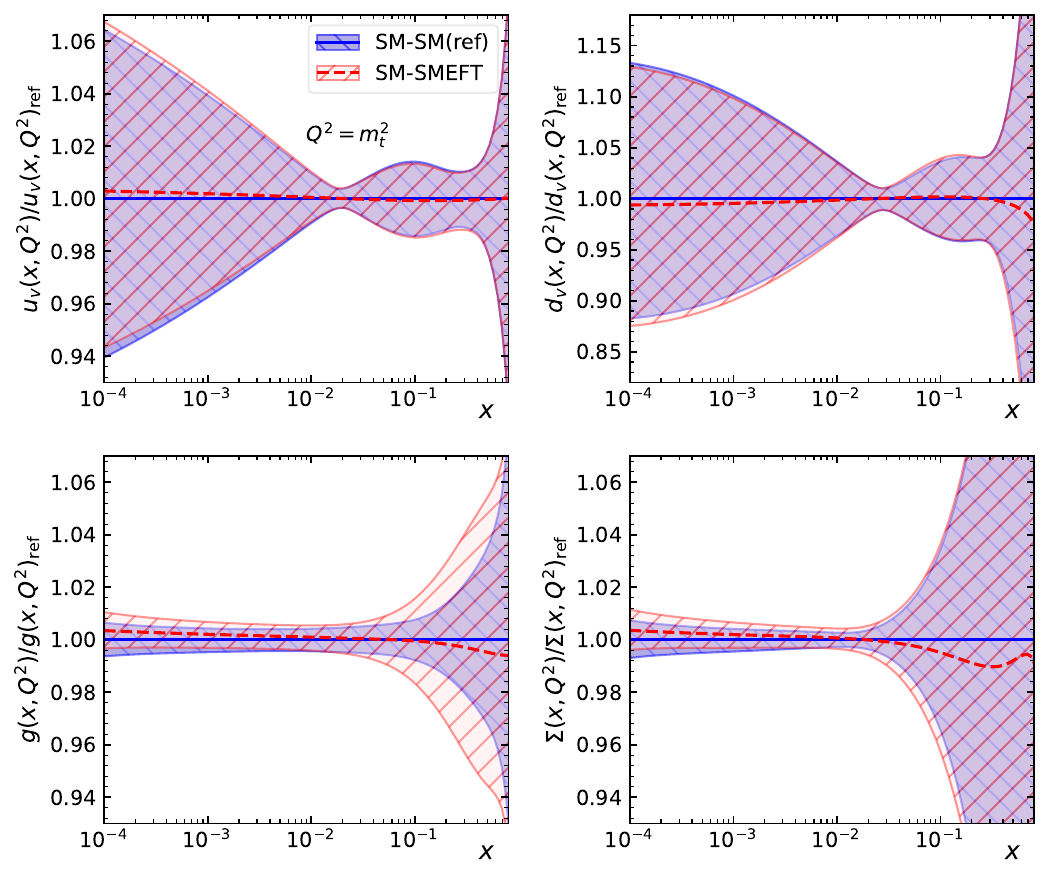}
 \caption{
    The valence, the gluon and the sea PDFs with their uncertainties, obtained in the SM-SM fit (blue shaded band) and the SM-SMEFT fits (red shaded band), normalised to the results of SM-SM fit. The PDFs are shown as a function of $x$ and the factorisation scale of $m_t$. The ratio of the results of the SM-SMEFT fit to those of SM-SM fit is presented by a dashed line.}
  \label{fig:PDFs-SMSM-SMSMEFT}
\end{figure}
In the SMEFT fit, both linear and quadratic corrections from SMEFT parameters are taken into account.
In Fig.~\ref{fig:PDFs-SMSM-SMSMEFT}, the PDFs obtained in the SM-SM and SM-SMEFT fits are compared.
The central values of the PDFs obtained in both fits almost coincide in the full $x$ range. 
Minor difference ($<5\%$) is observed for the quark PDFs in the large $x$ region 
where PDF uncertainties are large. The uncertainties in the gluon PDF increase, once the SMEFT coefficients are added as free parameters in the fit, 
indicating a reduction of constraining power of the $\ttbar$ data 
due to correlation between the gluon PDF and SMEFT parameters. In the same way, the uncertainty in $m_t$ increases. The results of the SM-SMEFT fit are consistent with the SM, with the obtained EFT coefficients consistent with zero, giving confidence that no fake signal of the new physics will be found in such a QCD analysis.

\begin{figure}[htbp]
\centering
\includegraphics[width=0.7\textwidth]{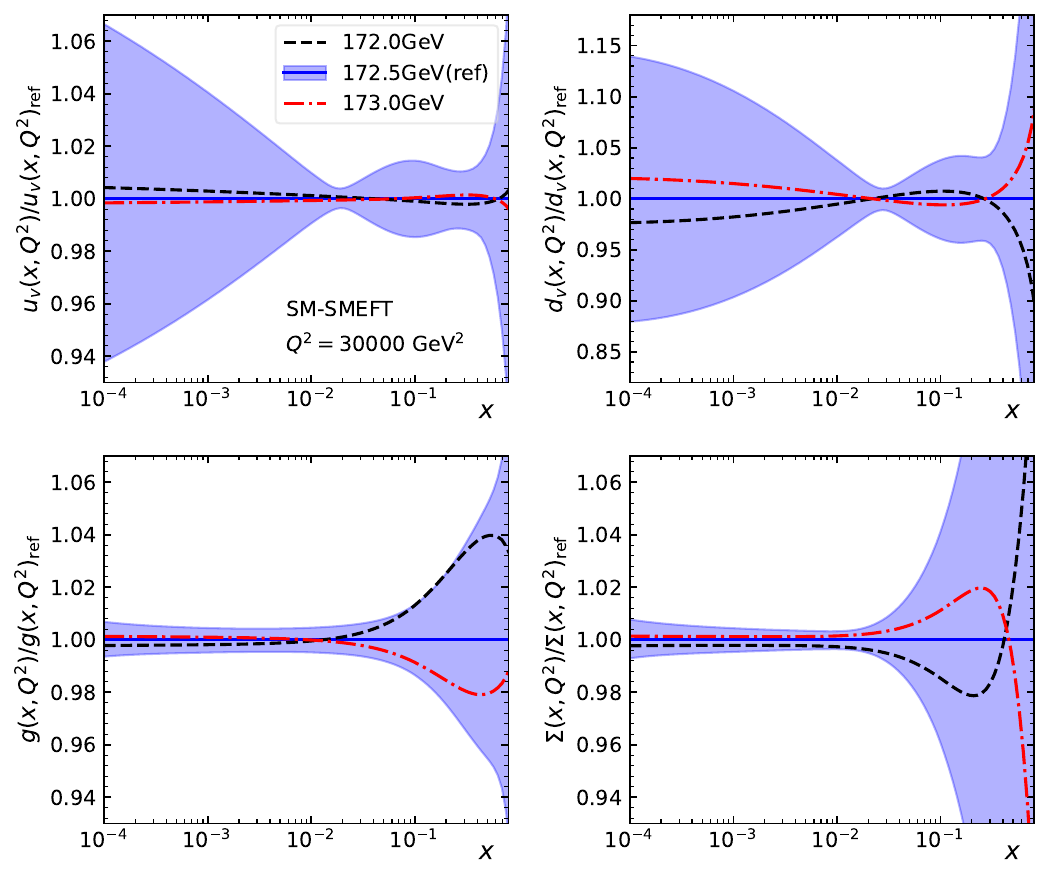}
  \caption{Valence, gluon and sea PDFs obtained in SM-SMEFT fits assuming different values of $m_t$ (lines of different styles), shown as functions of $x$ at factorisation scale of $30000 \GeV^2$. The 
  PDF uncertainties (shaded bands) for the choice of $m_t\!=\!172.5~{\rm GeV}$ are presented.
  }
  \label{fig:HERAPDF-fixed-mt}
\end{figure}
\begin{figure}[htbp]
\centering
\includegraphics[width=0.7\textwidth]{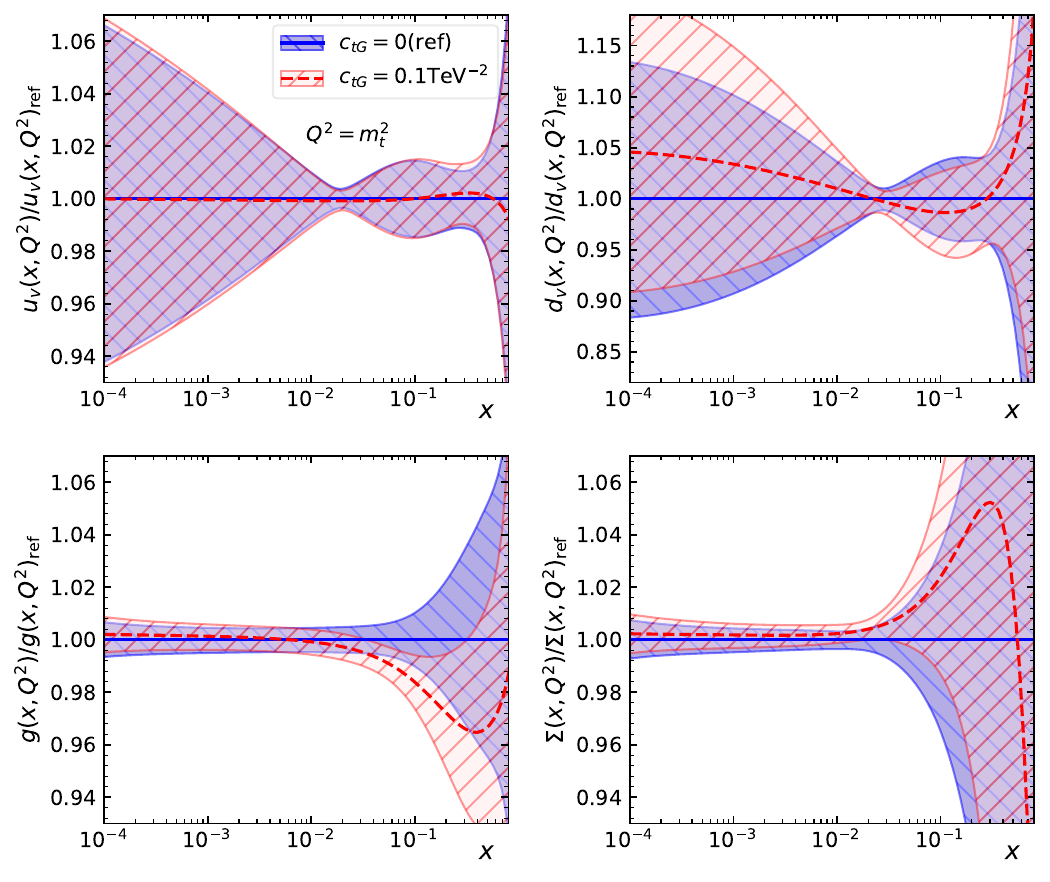}
  \caption{PDFs extracted from SM-SMEFT fits with fixed values of $\ctg=$
  $0$ and $0.1~\TeVmt$,
 at a scale of $Q^2=30000 \GeV^2$.
  }
  \label{fig:HERAPDF-fixed-ctg}
\end{figure}
The SM-SMEFT fit is also useful to study
the correlation between the gluon PDF and the values of $m_t$ or SMEFT parameters. The PDFs extracted in the SMEFT+PDF fits, where the alternative values for $m_t$=172.0, 172.5, or 173.0 GeV, are used, are presented in \Fig{fig:HERAPDF-fixed-mt}.
Smaller value of $m_t$ leads to reduction of the $t\bar t$ cross section close to the threshold, which would require larger contributions from the SMEFT coupling $\ctg$. 
To compensate for this effect at high $\mtt$, the gluon PDF is reduced at large $x$.
The PDFs extracted with alternative values of $\ctg=0.0$ and $0.1~\TeVmt$ are shown in \Fig{fig:HERAPDF-fixed-ctg}.
For the same reason as in the case of varying $m_t$, the increase of $\ctg$ leads to a reduction of the gluon PDFs at high $x$. 

\subsection{Fitting SMEFT pseudo-data}

Here, a scenario is considered, where new physics 
is present in the pseudo-data of $\ttbar$ production of a future HL-LHC run.
In particular, as detailed in \Sec{sec:data}, the pseudo-data is generated using the theoretical prediction assuming $\ctq\!=\!1.0~\TeVmt, \ctg\!=\!-0.1~\TeVmt$, considering both linear and quadratic EFT corrections. For simplicity, the other SMEFT couplings are set to zero. 
Again, two general scenarios are considered. In a {\it SMEFT-SM} fit, only PDFs and $m_t$ are fitted, attempting an SM-only interpretation of the data which do contain BSM effects. In the alternative {\it SMEFT-SMEFT} fit, the PDFs, $m_t$ and the EFT coefficients are extracted simultaneously, correctly considering the generated BSM effects, and fully accounting for the correlations between EFT and SM parameters. The results are summarised in the Table~\ref{tab:fit-res-SMEFT}  and are discussed in the following.

\begin{figure}[htbp]
\centering
 \includegraphics[width=0.47\textwidth]{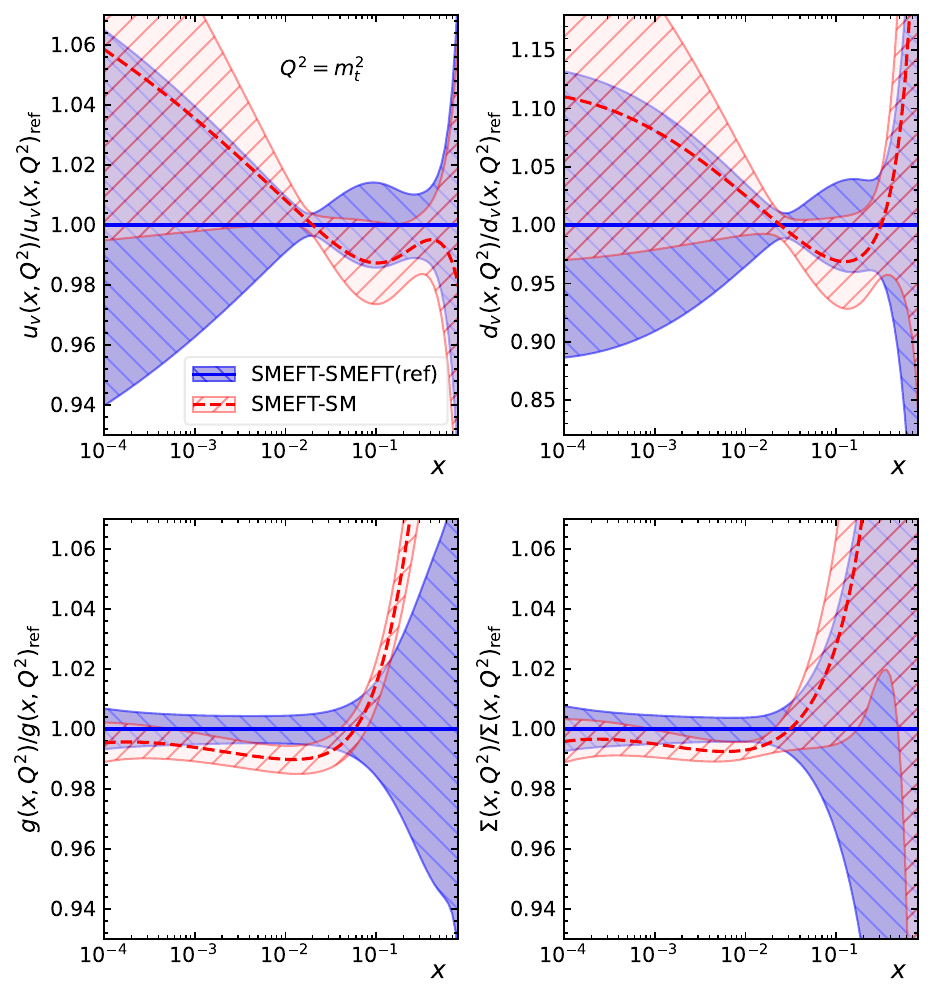}
  \quad
 \includegraphics[width=0.47\textwidth]{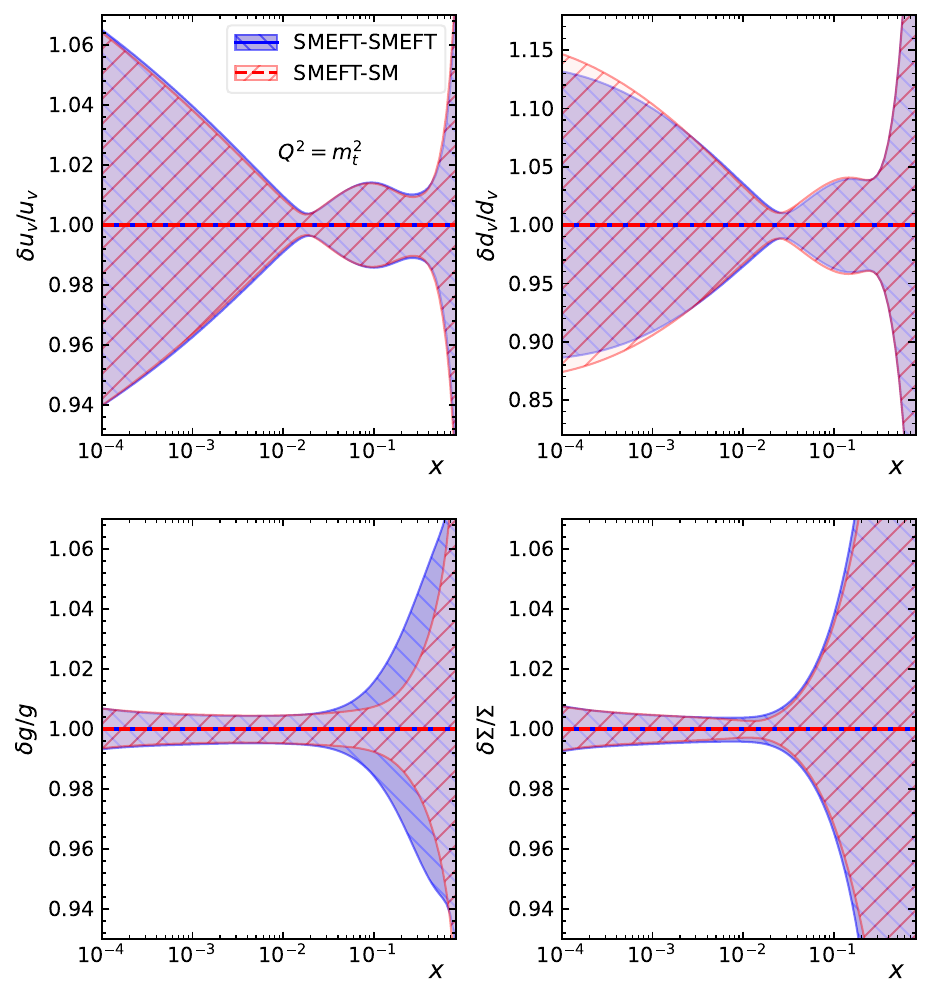}

	\caption{
    The valence, the gluon and the sea PDFs (left panel) and their relative uncertainties (right panel), obtained in the SMEFT-SMEFT fit (blue shaded band) and the SMEFT-SM fit (red shaded band), shown as a function of $x$ and the factorisation scale of $m_t$. The PDFs are normalised to the results of SMEFT-SMEFT fit and the ratio is presented by a dashed line.
}
  \label{fig:PDFs-SMEFTSMEFT-SMEFTSM}
\end{figure}

\begin{table}[h]
\centering
\setlength\tabcolsep{2.5pt}
  \begin{tabular}{l|c|c|c|c|c}
    \hline
     & $m_t$& $\ctg$ & $\ctq$ & $\ctu$ & $\ctd$\\ 
     & [GeV]& $[\TeVmt]$ & $[\TeVmt]$ & $[\TeVmt]$ & $[\TeVmt]$\\    
    \hline
    \small generated & $172.5$ & $-0.1$ & 1.0 & 0& 0 \\
    \hline
    {\small SMEFT-SMEFT (full)} & $172.50 \pm 0.37$ & $- 0.11 \pm 0.08$ & $1.00 \pm 0.25$ & $-0.01 \pm 0.37$ & $0.01 \pm 1.14$\\
    \hline
    {\small SMEFT-SMEFT (linear)} & $172.47 \pm 0.36$ & $-0.07\pm 0.39$ & $0.35\pm 12.22$
    & $-0.35 \pm 3.46$ & $8.36 \pm 53.34$\\
    \hline
    {\small SMEFT-SM} & $172.83 \pm 0.23$ & - & - & - & -\\\hline
    {\small fixed-PDF SMEFT} & $172.41\pm0.35$ & $-0.14\pm0.08$ & $0.93\pm0.50$ & $-0.01\pm1.05$ & $-0.09\pm1.69$\\ \hline
  \end{tabular}
  \caption{
  Results for $m_t$ and the Wilson coefficients
  obtained in the fits to HERA DIS and SMEFT $\mtt$ pseudo-data using different scenarios.
  The values in the first row correspond to the input parameters used to generate the pseudo-data. 
  }
  \label{tab:fit-res-SMEFT}  
\end{table}
\entry{SMEFT-SMEFT fit}
Here, the PDFs, $m_t$, and the SMEFT coefficients $\ctg,\ctq,\ctu,\ctd$ are extracted simultaneously in a full SMEFT interpretation of the pseudo-data generated with non-zero assumption on $\ctg$ and $\ctq$. The linear and quadratic corrections from SMEFT parameters are considered in the fit.
The results, labelled as {\it SMEFT-SMEFT} in the \Tab{tab:fit-res-SMEFT}, agree well with the truth values for all the inputs used to generate the pseudo-data, confirming that a comprehensive QCD analysis taking into account all the correlations among the SM and BSM parameters would return the correct result.
The PDFs extracted in the present SMEFT-SMEFT fit are shown in~\Fig{fig:PDFs-SMEFTSMEFT-SMEFTSM}.

\begin{figure}[htbp]
\centering
     \begin{subfigure}[b]{0.45\textwidth}
     \centering
     \includegraphics[width=\textwidth]{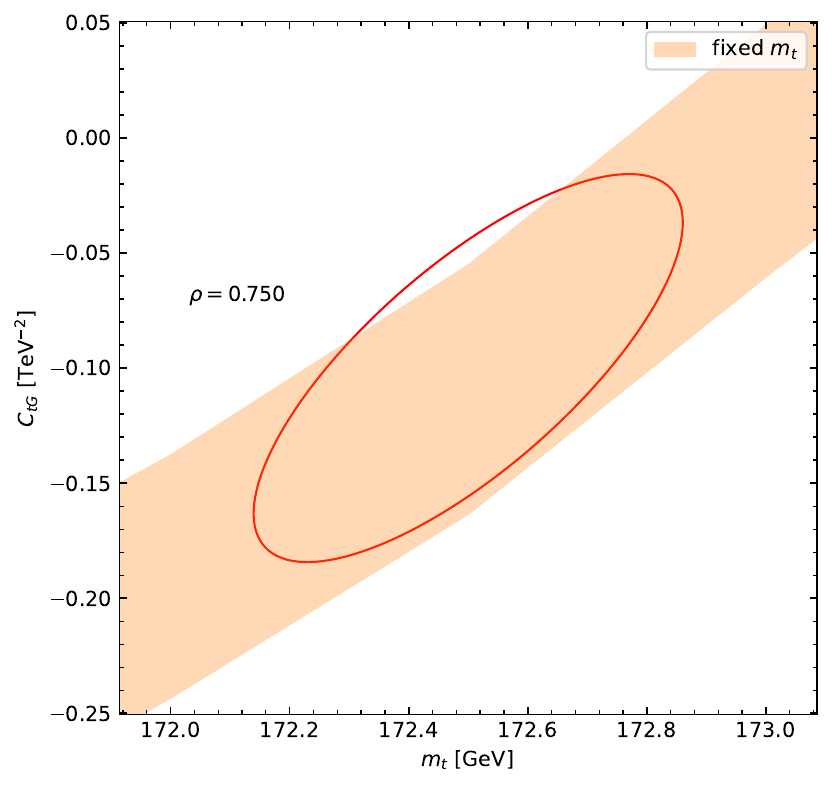}
     \caption{}
     \label{fig:contour-a}
     \end{subfigure}
    \begin{subfigure}[b]{0.45\textwidth}
    \centering  \includegraphics[width=\textwidth]{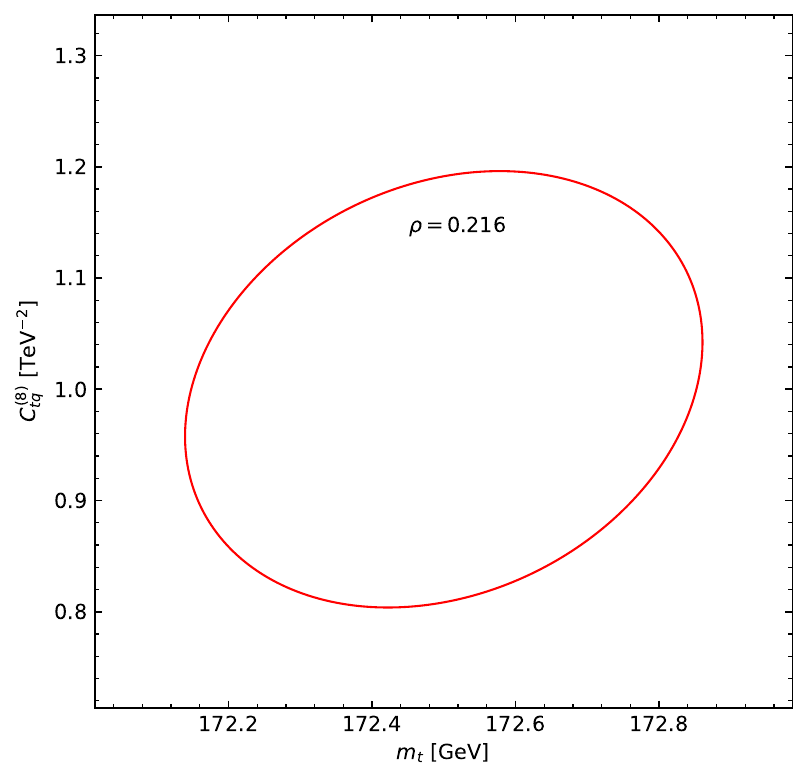}
    \caption{}
    \label{fig:contour-b}
    \end{subfigure}
    \begin{subfigure}[b]{0.45\textwidth}
    \centering \includegraphics[width=\textwidth]{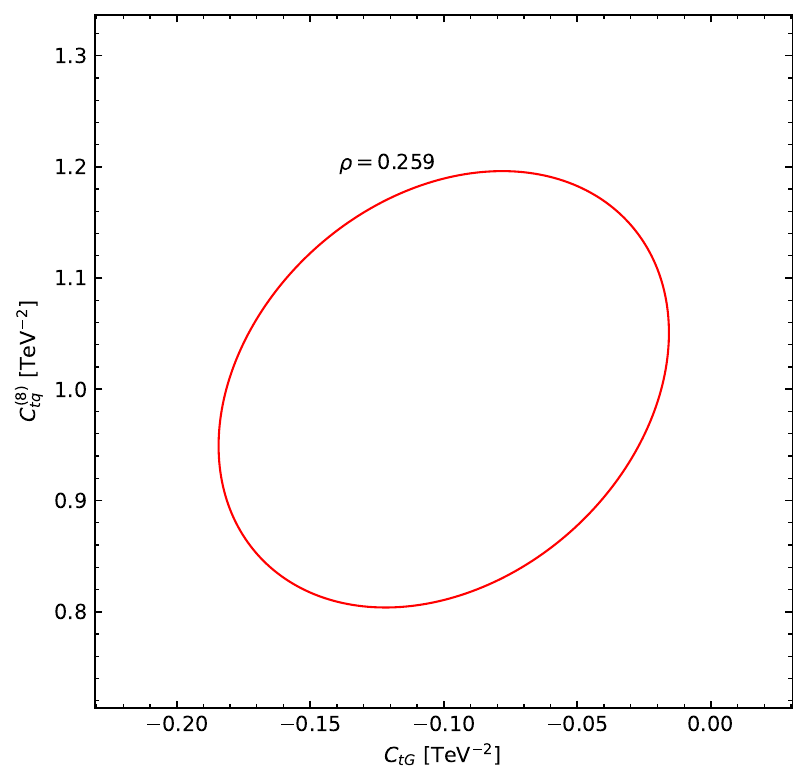}
    \caption{}
    \label{fig:contour-c}
    \end{subfigure}
    \caption{
    Correlations between $m_t$ and $\ctg$ (panel a), $m_t$ and $\ctq$ (panel b), and $\ctg$ and $\ctq$ (panel c). The dependence of $\ctg$ on an assumption on $m_t$ is presented by the shaded band.
    In each panel, the ellipse represents the $\Delta\chi^2=1$ contour based on the Hessian matrix. The correlation coefficients $\rho$ are also indicated.
    }
  \label{fig:contour-mt-SMEFT}
\end{figure}

Based on the result of the SMEFT-SMEFT fit, the correlation between the SM and EFT parameters is quantified in~\Fig{fig:contour-mt-SMEFT}. In~\Fig{fig:contour-a}, the correlations between $m_t$ and  $\ctg$ are presented. The value of $\ctg$ depends linearly on the assumed $m_t$. Once $m_t$ and $\ctg$ are considered as free parameters in the fit, both get constrained. Their correlation coefficient reaches $\rho_{m_t~\ctg}$=0.75. This strong positive correlation is expected, since the increase of $m_t$ reduces the cross sections close to the threshold $\mtt\simeq 2\mt$, which can be compensated by requiring a larger value of $\ctg$. 
The correlation of $m_t$ with $\ctq$, and of $\ctg$ with $\ctq$ is shown in \Fig{fig:contour-b} and Fig.~\ref{fig:contour-c},
respectively. 
Resulting correlation coefficients are smaller, with $\rho_{m_t~\ctq}$=0.216 and $\rho_{\ctg~\ctq}$=0.259.
That is expected because in contrast to $m_t$ and $\ctg$, the coefficient $\ctq$ has only an impact on the cross sections at high $\mtt$.
The correlation between $m_t$ and $\ctg$ could possibly be mitigated (and the parameters better constrained) by using multi-differential distributions in $t\bar t$ production, e.g. as a function of $m_{t\bar t}$ and rapidity of the $t\bar t$ pair, or rapidity difference of the top quark and anti-quark.
Of further advantage would be inclusion of the LHC measurements at different center of mass energies since these probe different composition of production channels via gluon fusion and quark annihilation.   

The importance of the quadratic SMEFT corrections is estimated in a variant of the SMEFT-SMEFT fit, labelled as {\it SMEFT-SMEFT (linear)}, where only linear  corrections are considered.
As a result, the values of 
$\ctq,\ctu$ and $\ctd$ are poorly constrained, as compared with the SMEFT-SMEFT full fit.
This can be attributed to the fact that at high $\mtt$, 
the quadratic SMEFT corrections are of the same size or even larger than the linear  corrections.

\entry{SMEFT-SM fit}
This study is performed in order to investigate potential bias in extracting the PDFs and $m_t$ from the data, which potentially incorporate effects of new physics. In this case, the SM fit, assuming all the SMEFT coefficients to be zero, is performed using the SMEFT pseudo-data.
In \Fig{fig:PDFs-SMEFTSMEFT-SMEFTSM}, the PDFs obtained from such a SMEFT-SM fit are shown and compared to the the correct SMEFT-SMEFT results.
While the valence distributions from both fits are still compatible within their uncertainties, the gluon PDF obtained in SMEFT-SM fit is significantly different from the correct SMEFT-SMEFT solution at large $x$. This is caused by the actual presence of BSM physics in the $\mtt$ pseudo-data, where the EFT $\ctq$ coupling contributes mostly to high invariant mass of the $t\bar t$ pairs, produced predominantly in gluon-gluon fusion, as illustrated in \Fig{fig:SMEFT-pred}.
At the same time, the uncertainties in the gluon PDF in the SMEFT-SM fit is underestimated at high $x$. The value of $m_t$ is also shifted due to disregarded correlation with $\ctg$ and the uncertainty is underestimated.
%


\entry{SMEFT using fixed PDF}
In an alternative analysis labelled as {\it fixed-PDF SMEFT}, 
we assume that no BSM effects are hidden in the data used for the PDF fit, while these are contained in the data used for the EFT interpretation.
Instead of a full QCD+SMEFT analysis of DIS and HL-LHC pseudo-data, an EFT interpretation of the latter is performed, as often done in a conventional BSM search.
{In the focus of this investigations are the uncertainties in both PDFs and EFT coefficients.}
First, we extract PDFs from the QCD fit to the HERA DIS data only, 
which is equivalent to the HERAPDF20\_NNLO\_EIG PDF set~\cite{Abramowicz:2015mha}.
This PDF set is then used in a SMEFT prediction to fit the values of $m_t$ and SMEFT parameters, using the  $\ttbar$ pseudo-data alone.
The PDF uncertainties are taken into account by treating all the PDF eigenvectors as nuisance parameters.
The results on $m_t$ and SMEFT parameters are similar to those obtained in the SMEFT-SMEFT fit, however with increased uncertainties in $\ctq,\ctu,\ctd$ coefficients.
Therefore, {
using only low energy data in the PDF fit may help to avoid contamination by New Physics.
However, the lack of high energy data leads to loose constraints on the resulting PDFs and the SMEFT parameters.
}

\section{Summary}
\label{sec:summary}
 Uncertainties in the knowledge of the proton structure
play an increasingly important role in the interpretation of the LHC measurements of Standard Model processes, requiring 
precise and unbiased determinations of SM or beyond SM (BSM) parameters 
to be performed simultaneously with a determination of PDFs.

This work presents a new module (so-called \reaction) for the \textsc{xFitter} program for simultaneous extraction of PDFs and various SM/BSM parameters. 
The new \reaction provides a fast parameterisation of the cross-sections dependence on SM/SMEFT+PDF, incorporating both linear and quadratic SM/BSM corrections and considering the full dependence of these corrections on the PDFs.
The \reaction can be also used for SM/SMEFT+PDF fits, or also for
SM/BSM parameters extraction without fitting PDFs.

The performance of the code is illustrated in an example exploring the sensitivity of $t\bar t$ production at the HL-LHC to selected operators in SM Effective Field Theory (SMEFT) framework. In particular, finite values for the EFT coefficients $\ctg$ and $\ctq$ are assumed\footnote{Alternatively, pseudo-data were generated with different values for $\ctg$ and $\ctq$, which didn't modify any of presented conclusions.} and the fits are performed in the Standard Model only and SMEFT scenarios. It is demonstrated that only careful consideration of the full correlations 
between PDFs, the value of $m_t$ and the EFT parameters, including SMEFT quadratic corrections, would lead to the correct interpretation of the data. The correlations among the investigated parameters are also quantified.

The presented \reaction is offered as a new basis for SM/BSM+PDF interpretation of the measurements within the \textsc{xFitter} framework, easily extendable to 
include arbitrary SM/SMEFT parameters and/or more datasets, according to the user case.


\section*{Acknowledgments}
This work is supported by the Helmholtz grant W2-W3/123 and by the Helmholtz-OCPC Postdoctoral Exchange Program under grant No. ZD2022004.
The work of JG is supported by the National Natural Science Foundation of China (NSFC) under Grant No. 12275173. 
The work of OZ is supported by the Philipp Schwartz-Initiative of the Alexander von Humboldt foundation.
XS would like to thank Toni Mäkelä and Marco Guzzi for helpful discussions.

\appendix
\section{Installation and usage of the \reaction}
\label{sec:install}
For general installation and basic usage of \textsc{xFitter}, the users will find it helpful to go through the documentation~\cite{xfittermain}.
The \textsc{xFitter} package is a powerful toolbox for extracting PDFs and $\alpha_s$ and incorporates various reactions for specific user cases.
The AFB reaction \cite{Anataichuk:2023elk} can be used to fit SMEFT parameters relevant for the 
forward-backward asymmetry $A_{\rm FB}$.
Quark contact interactions can be fitted using the CIJET reaction~\cite{CMS:2021yzl}.
It is also possible to fit $m_t$ to inclusive $\ttbar$ production rate data using the \textsc{xFitter} interface to HATHOR~\cite{Aliev:2010zk}.
The \reaction introduced in this work, on the other hand, 
aims to fit SM or EFT parameters 
for which a dedicated reaction is not yet available in the \textsc{xFitter}.
In the following, we will focus on the usage of the \textsc{xFitter} \reaction.

The \reaction has been integrated into \textsc{xFitter},
and is to be installed together with the main code. Users may also find it useful to adapt the examples in the \vb{examples/EFT} directory to their physics case.
For earlier versions of \textsc{xFitter}, the \reaction may also be added by:
\begin{enumerate}
    \item downloading the source code of the \reaction from the repository~\cite{xfitter1}    
    and adding the \verb"EFT" directory under
    \verb"/path_to_xfitter/xfitter_master/reactions"~;
    \item appending a new line 
\verb"add_subdirectory(reactions/EFT)" to \\
\verb"/path_to_xfitter/xfitter_master/CMakeLists.txt"~;
    \item recompiling \textsc{xFitter} by executing \\
\verb"cd /path_to_xfitter"\\
\verb"source setup.sh" \\
\verb"cd xfitter_master"\\ 
\verb"./make.sh install"~.
\end{enumerate}

In order to consider the full correlation between PDFs and the other free parameters in the fit,
the \reaction supports both \textsc{APPLgrid}~\cite{Carli:2010rw} and \textsc{PineAPPL}~\cite{Carrazza:2020gss} fast-interpolation grids.
If needed, the \textsc{PineAPPL} package has to be installed within \textsc{xFitter}, 
e.g. using the script from Ref.~\cite{xfitter2}
and setting \vb{skip\_pineappl=0}.
%

The \reaction has barely limit on integration of the measured observables (e.g. differential cross sections, or $A_{\tmop{FB}}$ etc.) and fitted parameters,
except for the key assumption that 
the contributions of these parameters to the theoretical predictions for these observables
can be well approximated by quadratic polynomials
\Eq{eq:quad-dep}, which we quote here for readers' convenience
\begin{eqnarray}
  \sigma^{(\alpha)} (\WC) & = & \sigma_0^{(\alpha)} + \sum_i c_i \sigma_i^{(\alpha)} + \sum_{i \leqslant j} c_i c_j
  \sigma_{i j}^{(\alpha)} \nonumber\\
  & \equiv & \sigma_0^{(\alpha)} + \sum_i c_i l_i^{(\alpha)} + \sum_i c_i^2 q_i^{(\alpha)} + \sum_{i < j} c_i c_j m_{i j}^{(\alpha)} \nonumber\\
  & = & \sigma_0^{(\alpha)} \left( 1 + \sum_i c_i K_i^{(\alpha)} 
  + \sum_{i\leqslant j} c_i c_j K_{i j}^{(\alpha)} \right) \,,
  \label{eq:quad-dep-app}
\end{eqnarray}
where $l_i\!\equiv\!\sigma_i,~q_i\!\equiv\!\sigma_{i i},~m_{i,j}\!\equiv\!\sigma_{i,j}$ denote the linear, quadratic, and mixed corrections, respectively.
Consideration of cubic or even higher power corrections is discussed below.

The expression in \Eq{eq:quad-dep-app} is the key input needed by the \reaction.
For example, the last expression of \Eq{eq:quad-dep-app} can be implemented in an \textsc{xFitter} data file using the \vb{TheorExpr} argument as
\begin{lstlisting}[language=yaml]
TermName   = 'SMNNLO',   'KEFT'
TermSource = 'PineAPPL', 'EFT'
TermInfo   = 
  'GridName=/path/to/PineAPPL_grid.pineappl',
  'ListEFTParam=deltamt,ctg,ctq8:FileName=/path/to/EFT_file.yaml
:xiF=1.0:xiR=1.0'
TheorExpr  = 'SMNNLO*KEFT'
\end{lstlisting}
where the theoretical prediction is the product of two terms, 
the SM prediction \vb{SMNNLO} and a $K$ factor \vb{KEFT}.  
The \vb{TermSource} designates the \textsc{xFitter} reactions used to calculate the corresponding terms, and 
\vb{TermInfo} provides variables passed to the reactions, separated by colons.
In this example, the \vb{SMNNLO} term is calculated by the \vb{PineAPPL} reaction,
with the path to the corresponding PineAPPL~\cite{Carrazza:2020gss} grid given by keyword \vb{GridName}.
The \vb{KEFT} term is a $K$ factor calculated with the \reaction,
whose ingredients we elaborate now.

\entryi{Mandatory arguments for the EFT Term}
There are two mandatory arguments for an EFT term.
The argument \vb{ListEFTParam} transmits to \textsc{xFitter} the list of the SM/EFT parameters to be fitted
\footnote{Note that the
PDF parameters or $\alpha_s (m_Z)$ do not need to be declared 
in the \vb{ListEFTParam} argument even if 
they are simultaneously fitted.}.
As with other fit parameters 
in \textsc{xFitter}, 
the initial values of the fit parameters $\{c_i\}$ shall be given in the \vb{parameters.yaml} file.
The corresponding BSM corrections, such as $l_i, q_{i}, m_i, K_i$ 
in \Eq{eq:quad-dep-app}, are assumed to be provided in the \vb{yaml} file,
whose format is described below.
The path to this \eftfile{} shall be given by the \vb{FileName} argument.

\entryi{Controlling the reaction output}
By default, the \reaction returns the ratio 
$\sigma(\WC)/\sigma(\WC\equal 0)$, 
which may correspond to the BSM factor $\sigma_{\rm SM+EFT}/\sigma_{\rm SM}$.
Here $\sigma(\WC)$ 
is the theoretical predictions corresponding to the current values of $\WC$, which are updated from one iteration to another during the fit.
%
This behavior can be changed by two optional arguments.
One is the \vb{AbsOutput} argument (\vb{False} by default), which determines if the output shall be divided by the central predictions $\sigma(\WC \equal 0)$,
and the other is the \vb{NoCentral} argument (\vb{False} by default), which toggles if the central contributions should present in the output.
Outputs corresponding to different combinations are summarized in \Tab{tab:option-abs-cen}.
\begin{table}[h]
  \centering
  \begin{tabular}{|l|l|l|}
    \hline
    \vb{AbsOutput} & \vb{NoCentral} & reaction output\\
    \hline
    False & False & $\sigmaSMBSM / \sigmaSM$\\
    \hline
    False & True & $\sigmaSMBSM / \sigmaSM-1$\\
    \hline
    True & False & $\sigmaSMBSM$\\
    \hline
    True & True & $\sigmaBSM$\\
    \hline
  \end{tabular}
  \caption{Outputs of the \reaction corresponding to different combinations of options.
  }
  \label{tab:option-abs-cen}
\end{table}

\entryi{Varying the scales}
A common practice to estimate the effect of missing higher-order correction is to vary the renormalisation and/or factorisation scales by a factor $\xi_{R,F}\equiv \mu_{R,F} / \mu_{R,F}^{(0)}$, 
with $\mu_{R}^{(0)}$ ($\mu_{F}^{(0)}$) being the nominal scale choice in the generation of the fast-interpolation grids.
This can be done with the \vb{xiR} and \vb{xiF} keywords in \reaction, 
(only applicable once the input linear/quadratic corrections are provided as APPLgrid or PineAPPL grids, instead of $K$ factors).
%

\entryi{The \eftfile{}}
As mentioned in \Sec{sec:framework},
the linear and quadratic corrections
$l_i, q_i, m_{i,j}$
in \Eq{eq:quad-dep-app}
have to be calculated outside \textsc{xFitter} 
and fed to \reaction via an \eftfile.
This \eftfile{} contains a list of entries, each corresponding to either of
\begin{itemize}
    \item a linear/quadratic correction ($l_i, q_i, m_{i,j}$), 
    \item a ratio ($K_i, K_{i j}$), or
    \item a theoretical prediction $\sigma(\WC)$ 
    (can be numbers or  fast-interpolation grids) 
    together with the values of $\WC$. 
\end{itemize}
Sufficient entries should be provided such that  
linear corrections $l_i$ for all fitted parameters $i$
(and also quadratic corrections $q_i, m_{i,j}$ if available)
can be determined by solving a system of linear equations.
Quadratic corrections $q_i, m_{i,j}$ that can not be extracted 
due to lack of necessary inputs are assumed to be zero.

As an example, an \eftfile{} of the form
\begin{lstlisting}[language=yaml]
# The EFT YAML file for fitting ctg
SM_NLO: # name of the entry is almost arbitrary
  type: C # Central predictions (sigma_0 in Eq.(A1))
  format: PineAPPL # xsec are PineAPPL tables
  xsec: [ /path/to/SM_NLO.pineappl ]
Linear_ctg: # This starts a new entry
  type: L # predictions up to Linear corrections
  param: ctg # name of the parameter (in ListEFTParam)
  param_value: 20.0 # value of ctg used to generate the grid
  format: PineAPPL
  xsec: [ /path/to/ctg1.pineappl ] # SM_NLO + 20.0*l_ctg
Quadratic_ctg: # SM_NLO + 40.0*l_ctg + 40.0^2*q_ctg
  type: Q # predictions up to Quadratic corrections
  param: ctg
  param_value: 40.0
  format: PineAPPL
  xsec: [ /path/to/ctg2.pineappl ]
\end{lstlisting}
can be used to fit the parameter \vb{ctg}.
This \eftfile{} has three entries, and the three involved grids provide the theoretical predictions (suppressing the units $\TeVmt$ for simplicity),
\begin{eqnarray}
    \sigma(\ctg=0) &\equiv& \sigma_0\nonumber\\
    \sigma_{\rm lin.}(\ctg=20) &=&  \sigma_0 + 20\cdot l_\ctg \nonumber\\
    \sigma(\ctg=40) &=&  \sigma_0 + 40 \cdot l_\ctg  + 40^2 \cdot q_\ctg
\label{eq:entry-ctg}
\end{eqnarray}
from which the \reaction extracts $l_\ctg$ and $q_\ctg$ 
and calculates the observable $\sigma(\ctg)$ for any value of $\ctg$
during the fit.
For each entry in the above \eftfile, the theoretical predictions are given by the \vb{xsec} node.
The \vb{type}, \vb{param}, and \vb{param\_value} nodes work together to 
define the interpretation of \vb{xsec}, as is summarized in 
\Tab{tab:eft-file-type}.

\begin{table}[h]
  \centering
  \begin{tabular}{|c|c|c|l|}
    \hline
    \vb{type} & \vb{param} & \vb{param\_value} & \vb{xsec} \\
    \hline
    C & - & - & $\sigma_0$\\
    \hline
    l & $i$ & - & $l_i \equiv \sigma_i$\\
    \hline
    q & $i$ & - & $q_i \equiv \sigma_{i i}$\\
    \hline
    m & $[i, j]$ & - & $m_{i j} \equiv \sigma_{i j}$\\
    \hline
    L & $i$ & $c_i$ & $\sigma_0 + c_i \sigma_i$\\
    \hline
    Q & $i$ & $c_i$ & $\sigma_0 + c_i l_i + c_i^2 q_{i}$\\
    \hline
    M & $[i, j]$ & $[c_i, c_j]$ & $\sigma_0 + c_i l_i + c_j l_j +
    c_i^2 q_{i} + c_j^2 q_{j} + c_i c_j m_{i j}$\\ \hline
  \end{tabular}
  \caption{
  Possible \vb{type}'s of the \eftfile{} entries,
  the needed \vb{param}, \vb{param\_value} inputs,
  and the corresponding theoretical predictions (\vb{xsec}), 
  using the notation of \Eq{eq:quad-dep-app}.
  The theoretical predictions should be further divided by $\sigma_0$ if \vb{format} node is set to \vb{ratio}. 
  }
  \label{tab:eft-file-type}
\end{table}

The above \eftfile{} can be extended for extraction of $\ctqb$ and the top quark mass
$\mt$ as follows.
\begin{lstlisting}[language=yaml]
# appended to the former EFT YAML file, to fit also ctq8 and mt
Quadratic_ctq8_1: # SM_NLO + 40*l_ctq8 + 40^2*q_ctq8
  type: Q
  param: ctq8
  param_value: 40.0
  #... inputs similar to Quadratic_ctg
Quadratic_ctq8_2: # SM_NLO + 30*l_ctq8 + 30^2*q_ctq8
  type: Q
  #... inputs similar to Quadratic_ctg
ctg_ctq8: # theoretical predictions for ctg=20, ctq8=40
  type: M # include the Mixing between ctg and ctq8
  param: [ctg, ctq8] # both ctg and ctq8 are non-zero
  param_value: [20.0, 40.0] # ctg=20, ctq8=40
  format: PineAPPL
  xsec: [ /path/to/ctg_ctq8.pineappl ]
Kfactor_mt_linear: # l_deltamt / Central
  type: l
  param: deltamt
  format: ratio # xsec is an array of ratios
  xsec: [-0.10349132, -0.02400355, -0.01317847, ... ] 
\end{lstlisting}
This way, $l_\ctqb,~ q_\ctqb,~ m_{\ctg, \ctqb}$, and $\l_{\mt}$ can be fitted.
The information 
about the quadratic corrections to $m_t$
and the mixing term between $m_t$ and the two SMEFT parameters, 
however, are missing and thus 
quadratic corrections
$q_{\delta m_t}$, $m_{\delta m_t,\ctg}$, $m_{\delta m_t, \ctqb}$ will not be included in calculating the theoretical predictions.

We have shown in \Fig{fig:BSM-KFactor} that PDF dependence 
does not always cancel out completely in the BSM $K$ factors,
therefore the fast-interpolation grids are preferred to 
fully incorporate the correlation between the PDFs and SM/SMEFT parameters.
{
  However, when the number of EFT parameters to be fitted $N$ is large,
  the number of involved fast-interpolation grids is also large ($O(N^2)$ for quadratic
  EFT fits or $O(N)$ for linear EFT fits).
  As a result, the calculation of theoretical prediction can be slow.
  %
}
For \textsc{xFitter} \reaction, theoretical predictions can be either numeric (obtained for a particular choice of the PDF set) 
or fast-interpolation grids (essentially disentangled from the PDFs).
This is indicated by the \vb{format} node, which can be 
\vb{PineAPPL} (\vb{APPLgrid})  for the grids in \textsc{PineAPPL} (\textsc{APPLgrid}) format, or by 
\vb{xsection} for numerical values ($\sigma(\WC), \sigma_i$ or $\sigma_{i j}$),
or \vb{ratio} for ratios $K_i = \sigma_i / \sigma_0$,  $K_{i j} = \sigma_{i j} / \sigma_0$, as is the case of \vb{deltamt} in the above \eftfile.
\entryi{Higher power corrections} To a limited extent, the cubic or even higher power corrections can be considered.
As an example, corrections proportional to $(\delta \mt)^3$ and $\ctg\cdot(\delta \mt)^3$ can be provided via the following two entries 
with \vb{monomial} type in the \eftfile:
\begin{lstlisting}[language=yaml]
# appended to the EFT YAML file above
mt_cubic: # contribution proportional to deltamt^3
  type: monomial
  format: ratio
  param: [deltamt]
  power: [3]
  xsec:  [0.0001, ...]
mt3_ctg: # contribution proportional to deltamt^3 * ctg
  type: monomial
  format: ratio
  param: [deltamt, ctg]
  power: [3, 1]
  xsec:  [0.0002, ...]
\end{lstlisting}
For the cubic or higher power corrections, the current version of \reaction can only deal with corrections from monomials.
Unlike \vb{L, Q, M} entries in \Tab{tab:eft-file-type} or \Eq{eq:entry-ctg}, these corrections can not be provided as linear combination of 
monomials.
 \bibliography{main-EPJC.bbl}

\end{document}